\documentclass{osa-article}

\journal{oe}

\usepackage{xcolor}
\usepackage{comment}
\usepackage{amsmath}
\usepackage{siunitx}
\usepackage{multirow}
\usepackage{subfig}
\usepackage{booktabs}
\usepackage{algorithm}
\usepackage{algorithmic}

\graphicspath{{figures/}}

\def\bs{\boldsymbol}
\def\di{\mathrm{d}}
\newcommand{\rdb}[1]{\left( #1 \right)}

\begin{document}

\title{Onsite Non-Line-of-Sight Imaging via Online Calibrations}

\author{Zhengqing Pan\authormark{1,2,3,6}, Ruiqian Li\authormark{1,6}, Tian Gao\authormark{4}, Zi Wang\authormark{1}, Ping Liu\authormark{1}, Siyuan Shen\authormark{1}, Tao Wu\authormark{1}, Jingyi Yu\authormark{1,5,*}, Shiying Li\authormark{1,5,*}}

\address{\authormark{1} School of Information Science and Technology, ShanghaiTech University, Shanghai 201210, China\\
\authormark{2} Shanghai Institute of Microsystem and Information Technology, Chinese Academy of Sciences, Shanghai 200050, China\\
\authormark{3} University of Chinese Academy of Sciences, Beijing 100049, China\\
\authormark{4} School of Physical Science and Technology, ShanghaiTech University, Shanghai 201210, China\\
\authormark{5} Shanghai Engineering Research Center of Intelligent Vision and Imaging, ShanghaiTech University, Shanghai 201210, China\\
\authormark{6} These authors contributed equally
}
\email{\authormark{*} yujingyi@shanghaitech.edu.cn}
\email{\authormark{*} lishy1@shanghaitech.edu.cn}

\begin{abstract}
There has been an increasing interest in deploying non-line-of-sight (NLOS) imaging systems for recovering objects behind an obstacle. Existing solutions generally pre-calibrate the system before scanning the hidden objects. Onsite adjustments of the occluder, object and scanning pattern require re-calibration. We present an online calibration technique that directly decouples the acquired transients at onsite scanning into the LOS and hidden components. We use the former to directly (re)calibrate the system upon changes of scene/obstacle configurations, scanning regions, and scanning patterns whereas the latter for hidden object recovery via spatial, frequency or learning based techniques. Our technique avoids using auxiliary calibration apparatus such as mirrors or checkerboards and supports both laboratory validations and real-world deployments.
\end{abstract}

\section{Introduction}

\begin{figure}[t]
    \centering
    \subfloat[Conventional scenario of NLOS imaging]{
        \begin{minipage}{.49\linewidth}
            \centering
            \includegraphics[width=.99\linewidth]{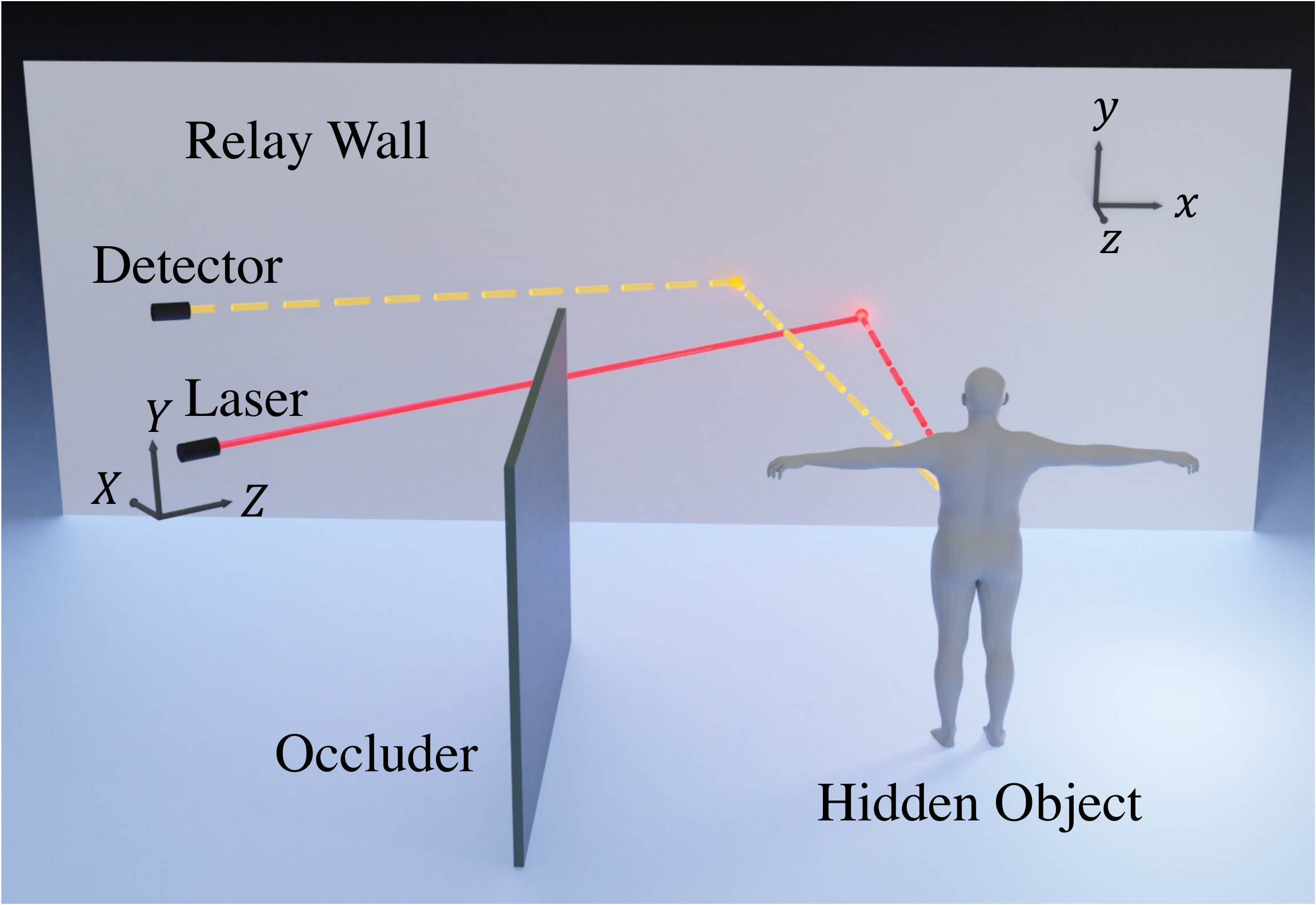}
        \end{minipage}
    \label{subfig:scenerio}}
    \subfloat[Top-view of setting]{
        \begin{minipage}{.49\linewidth}
            \centering
            \includegraphics[width=.99\linewidth]{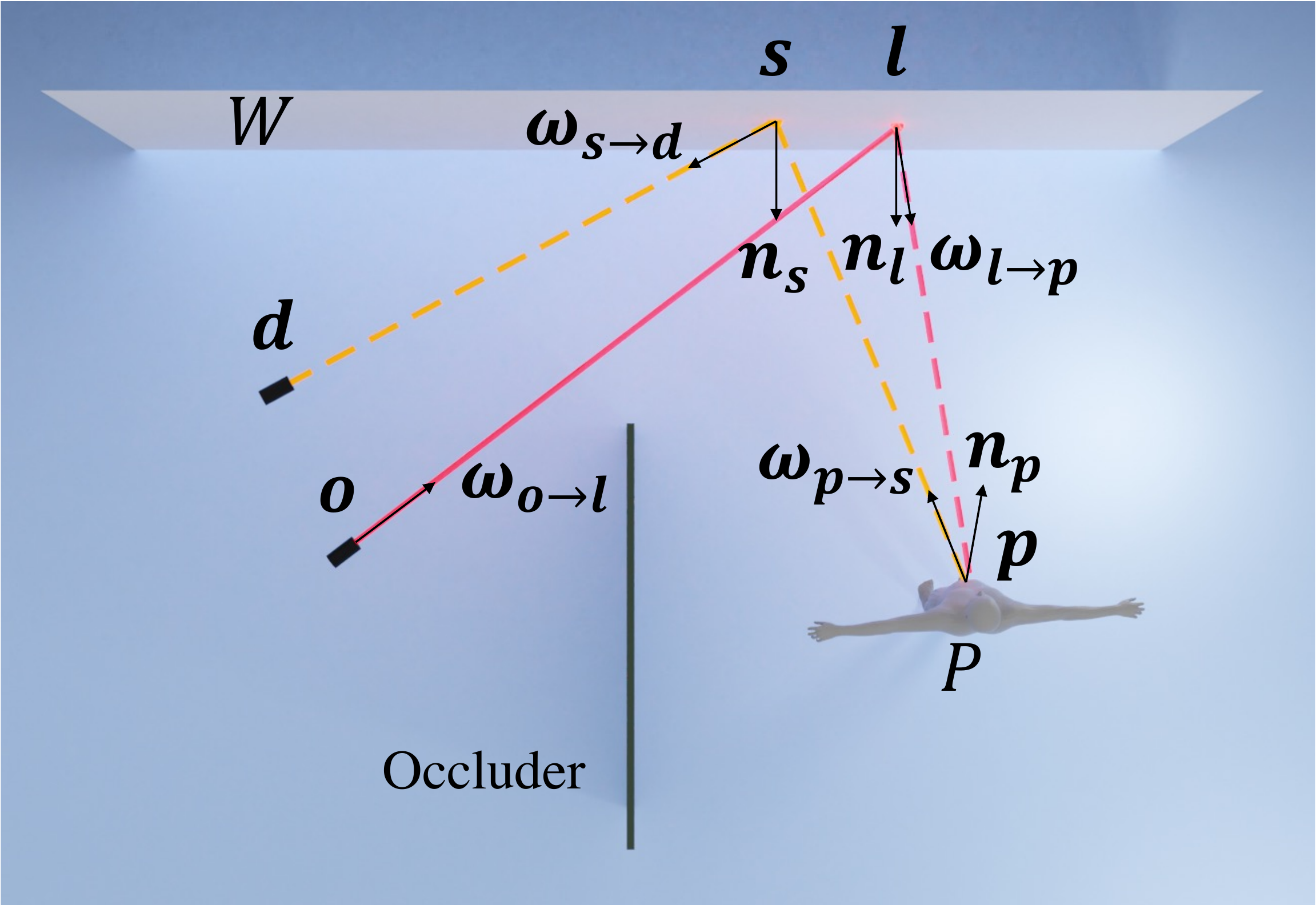}
        \end{minipage}
    \label{subfig:scenarioTop}}
    \caption{A conventional setting of the capture system for NLOS imaging: (a) conventional scenario with an ultra-fast pulsed laser and a time-resolved detector; (b) top-view of the setting.}
    \label{fig:scene}
\end{figure}

Non-line-of-sight (NLOS) imaging aims at recovering objects outside the direct line of sight of a sensor~\cite{2020Faccio, 2021Geng}. Most active NLOS capture systems exploit an ultra-fast pulsed laser beam that can be controlled to direct toward a relay surface (e.g., a wall). A companion time-resolved detector then collects the arrival time and number of photons that return after the first and third of three bounces: off the relay surface, off the hidden objects, and back off the relay surface. Fig.~\ref{fig:scene} illustrates a conventional setting of the NLOS capture system. The first bounce corresponds to direct reflection and enables us to recover the shape and albedo of the relay surface. By removing or gating the photons from the first bounce, photons from the third bounce are employed to reconstruct~\cite{2012Gupta, 2015Mauro, 2018LCT, 2019FK, 2019Liu, 2020ECCV, 2021Feng, 2021PNAS} and localize~\cite{2017Chan, 2018LCT, 2019FK, 2021PNAS, 2021Metzler} hidden objects in the NLOS scene. Potential applications are numerous, including autonomous driving, remote sensing, and biomedical imaging.

Substantial efforts have been made to both improve the hardware of the capture system \cite{2012Velten, 2015Mauro, 2018LCT, 2021PNAS, 2020Ceiling} and provide better NLOS reconstruction\cite{2019FK, 2019Liu, 2019Xin, 2021Ye, 2021Shen}. For the former, the efforts have been focused on using more accurate and affordable detectors and lasers. Streak cameras pioneered NLOS imaging ~\cite{2012Gupta, 2012Velten, 2014Wu, 2021Feng} by providing precise temporal resolution, e.g., down to 2 picoseconds (ps) or \SI{0.6}{mm}~\cite{2012Velten}. Such cameras, however, are overwhelmingly expensive and at the same time difficult to calibrate under nonlinear temporal-spatial transforms. Photonic mixer devices (PMDs) are compact and less expensive but can only provide low temporal resolution on the order of nanoseconds\cite{2013Heide}. In recent years, single photon avalanche diodes (SPADs) have served as an affordable and convenient alternative~\cite{2015Mauro, 2016Gariepy, 2018LCT, 2019FK, 2019Liu, 2020Ceiling, 2021PNAS}. A single-pixel SPAD with a time-correlated single photon counting (TCSPC) module produces a histogram of photon counts versus time bins of \SI{4}{ps} at a detection point. On the laser front, a picosecond or femtosecond pulsed laser is used to illuminate a spot on the relay surface, and to trigger and synchronize pulse signals with the detector so as to record the arrival times of returning photons. The properties of the laser, e.g., average power and repetition rate, affect acquisition time and measurements. 

Despite all these advances, constructing a prototype system even for validation still requires elaborate calibrations to acquire accurate transients and other information such as the 3D coordinates of scanning points. In a typical NLOS camera system, a uniform grid of $N \times N$ evenly spaced points is adopted to fit most reconstruction algorithms. An emerging trend, however, is to use non-uniform sampling, in both LOS\cite{2020NeRF} and NLOS\cite{2021Shen} imaging where the capture system would require re-calibration. In real-world deployment, it is also common to adjust the position and orientation of the occluder and the hidden object for validations where re-calibration needs to be conducted.

By far nearly all existing NLOS calibration schemes require auxiliary apparatus. For example, a checkerboard or regular point grid coupled with a visible camera\cite{2015Mauro,2019Ahn} can be used to acquire the internal and external parameters of the camera or the 3D coordinates of the grid point. Alternatives include using mirrors\cite{2020Klein} to establish correspondences of the laser point and the detection point for simultaneously estimating the mirror plane and the relay wall. When the scanning point is re-positioned, or the position of the relay surface changes relative to the data acquisition system, these methods need re-calibration. This is infeasible for onsite deployment for the process is both labor intensive and time consuming. 

We present an online calibration technique that directly decouples the acquired transients at onsite scanning into the LOS and hidden components. The former, we call \textit{Gamma} map, contains information of both geometry and albedo of the relay surface and detection points. We show how to use the \textit{Gamma} map to directly (re)calibrate the system upon changes of scene/obstacle configurations, scanning regions, and scanning patterns, as well as for transient normalization. This also provides a useful preview to allow dynamic adjustment: we can scan a small number of detection points and preview the calibration results of the scanning galvanometer and the relay surface to determine if the current scanning setup is acceptable. 

Furthermore, since our technique supports both uniform and non-uniform scanning patterns, we integrate existing NLOS reconstruction techniques for processing the hidden component based on spatial, frequency or learning based techniques. Many fast Fourier transform (FFT)-based algorithms, e.g., LCT\cite{2018LCT}, FK\cite{2019FK}, and PF\cite{2019Liu}, require transients in a regular grid as input whereas learning-based methods, such as neural transient fields (NeTF)~\cite{2021Shen}, allow more flexible sampling patterns but require significantly longer computational time, e.g., to train the network. We present a tailored reconstruction scheme suitable for arbitrary sampling patterns. This scheme employs the gradient descent method to iteratively reconstruct NLOS scenes. By supporting online calibration and onsite reconstruction assessment, our solution supports more feasible laboratory validations and real-world deployments and will be made open-source to the community.

\section{Related work}

NLOS imaging have used a variety of sensing technologies at the convergence of physics, optics, electronics, and signal processing. Optical detectors, including optical interferometry~\cite{2015Ioannis}, PMDs~\cite{2013Heide}, SPADs~\cite{2015Mauro, 2016Gariepy, 2018LCT}, intensified CCD cameras (iCCDs)~\cite{2014Martin}, streak cameras~\cite{2011Kirmani, 2012Gupta, 2012Velten, 2014Wu, 2021Feng}, or even non-optical devices, e.g., acoustic~\cite{2019Lindell}, thermal~\cite{2019Maeda}, and radar~\cite{2020Radar}, are exploited to collect transients of hidden objects. The temporal resolution of a detector is one of the key parameters for NLOS capture systems. Among the incoherent optical devices, PMDs and iCCDs are compact and low-cost and offer temporal resolution on the order of nanoseconds. Streak cameras and SPADs have high resolution of several picoseconds, but the former are bulky and incompatible with most capture systems in practice. For more details of time-resolved detectors, we refer readers to recent reviews~\cite{2017Jarabo, 2020Faccio, 2021Geng}. \\[-1.5ex] 

\textit{SPAD-based NLOS imaging.} Buttafava et al.\cite{2015Mauro} built the first SPAD-based NLOS capture system, which is a conventional setting, using a femtosecond (fs) laser that generates \SI{250}{fs} pulses at a wavelength of \SI{515}{nm} with a \SI{55}{MHz} repetition rate and an average power of \SI{50}{mW}. SPADs are reverse-biased photodiodes in Geiger mode, which distinguishes them from linear avalanche photodiodes (APDs). In particular, a SPAD restricts one incident photon to trigger an avalanche event and remains blind for a hold-off period, and then detects the next incoming photon. Several main factors, including sensitivity, dark count rate, temporal jitter, and hold-off time, define the signal-to-noise ratio and the temporal resolution and are related to the physical configuration and the material properties of the SPAD. For instance, silicon-based SPADs cover the visible spectrum, e.g., \SI{400}-\SI{800}{nm}, with full width at half maximum (FWHM) of tens of ps~\cite{2018LCT, 2019FK, 2019Liu}, whereas InGaAs/InP-based SPADs cover the infrared spectrum with FWHM of approximately \SI{200}{ps}~\cite{2021Ye, 2017Chan, 2021PNAS}. Wu et al.\cite{2021PNAS} constructed a long-range NLOS capture system, over \SI{1.43}{km}, exploiting InGaAs/InP SPADs at \SI{1550}{nm} and a fiber laser with \SI{500}{ps} pulses and \SI{300}{mW} average power.

We opt for a silicon-based single-pixel SPAD with a fast-gating mode, which can switch off the direct light paths between the relay surface and the SPAD. A 1D or 2D SPAD array simultaneously records transients of many pixels, but is often composed of single SPADs that have a smaller active area (of $\SI{6.95}{\um} \times \SI{6.95}{\um}$) and lower temporal resolution (hundreds of ps) due to the complicated fabrication~\cite{2016Gariepy, 2017Chan, 2021Pei}. Our SPAD has an active area of $\SI{50}{\um} \times \SI{50}{\um}$ and \SI{50}{ps} temporal jitter. The single SPAD in Stanford's NLOS capture system has a $\SI{100}{\um} \times \SI{100}{\um}$ active area and the laser is \SI{670}{nm} wavelength with \SI{30}{ps} pulses, \SI{10}{MHz} repetition rate, and \SI{0.11}{mW} average power\cite{2018LCT, 2019Heide}. Their improved version uses the fast-gated SPAD and a laser at \SI{532}{nm} with \SI{35}{ps} pulses and an average power of stronger than \SI{1}{W} \cite{2019FK}. They constructed several publicly available datasets of various types of NLOS objects: retro-reflective, specular, and diffuse. Using similar hardware devices, more NLOS capture systems have been built, and have achieved a large number of returning photons from the hidden objects within short acquisition\cite{2019Liu, 2019Ahn, 2019Xin}. These NLOS capture systems, as well as ours, are all of confocal setting with a beam splitter that locates the laser and the SPAD coaxially. In contrast, our laser is lower-cost, at a \SI{670}{nm} wavelength with \SI{50}{ps} pulses, \SI{40}{MHz} repetition rate, and \SI{2}{mW} average power, and produces up to 60 photons per pulse, which is sufficient for recovering many NLOS scenes. \\[-1.5ex]

\textit{Calibration schemes.} In NLOS imaging, positions and distances of the hardware devices need to be under control to record precise transients of hidden objects. Klein et al.\cite{2020Klein} proposed a calibration scheme with mirrors as target objects placed at different positions in an NLOS scene. Since the laser is directed by a 2D galvanometer, digital cameras are useful tools to take pictures of laser spots on the relay surface and to determine the 3D coordinates of the spots using computer vision techniques\cite{2015Mauro, 2019FK, 2019Ahn}. By scanning a pre-defined uniform grid with equidistant points, the position and orientation of the relay surface can also be estimated. However, these calibration schemes require specific target objects, e.g., mirrors or a checkerboard, and the digital cameras to be known or pre-calibrated. In contrast, our calibration technique needs no additional target objects or a digital camera and can operate online the iterations of calibration procedures. In particular, most previous studies attempt to determine the coordinates of scanning points or their mapping to input voltages of the galvanometer, resulting in inflexible scanning point selection and re-calibration while adjusting the relay surface or the galvanometer. Here we describe a detailed galvanometer calibration that allows us to arbitrarily select scanning points and to determine scanning patterns by manually or automatically controlling the galvanometer. Our calibration scheme enables users to calibrate the relay surface along with the galvanometer, and to calibrate these two separately for more precise effects.  

\section{NLOS imaging models}

NLOS imaging are constructed in accordance with an imaging formulation that models the physics of light traveling from a laser to a detector via LOS and NLOS objects. Fig.~\ref{fig:scene}\subref{subfig:scenarioTop} illustrates a top-view of the conventional setting. The pulsed laser beam $\mathbf{o}$ illuminates a spot $\mathbf{l}$ on a relay surface $W$. After the light scatters off the spot, some photons bounce off from a point $\mathbf{p}$ of the NLOS object $P$ and travel back onto a patch $\mathbf{s}$ of the relay surface. The detector $\mathbf{d}$ collects a number of photons from the patch at a time instant $t$. Based on the physics of light transport \cite{2006Dutre}, we define the image formation model as:
       
            \begin{equation}
                \begin{aligned}
                    \tau(t;\mathbf{o},\mathbf{l},\mathbf{s},\mathbf{d}) 
                    = \frac{N_{\mathbf{o}}}{c} \rdb{\frac{\rho}{\pi}}^{2} A_{\mathbf{d}} A_{\mathbf{s}} \frac{\rdb{\bs{\omega}_{\mathbf{s}\to\mathbf{d}} \cdot \mathbf{n}_{\mathbf{s}}} }{|\mathbf{s}-\mathbf{d}|^{2}}
                    \int_{P} \delta\rdb{|\mathbf{l}-\mathbf{o}|+|\mathbf{p}-\mathbf{l}|+|\mathbf{s}-\mathbf{p}|+|\mathbf{d}-\mathbf{s}| - ct} &\\ 
                    \cdot 
                    \upsilon(\mathbf{p};\mathbf{l},\mathbf{s})
                    f(\mathbf{p};\bs{\omega}_{\mathbf{l}\to\mathbf{p}},\bs{\omega}_{\mathbf{p}\to\mathbf{s}})
                    \frac{\rdb{\bs{\omega}_{\mathbf{p}\to\mathbf{l}} \cdot \mathbf{n}_{\mathbf{p}}} \rdb{\bs{\omega}_{\mathbf{l}\to\mathbf{p}} \cdot \mathbf{n}_{\mathbf{l}}} \rdb{\bs{\omega}_{\mathbf{s}\to\mathbf{p}} \cdot \mathbf{n}_{\mathbf{s}}} \rdb{\bs{\omega}_{\mathbf{p}\to\mathbf{s}} \cdot \mathbf{n}_{\mathbf{p}}}}{|\mathbf{l}-\mathbf{p}|^{2}|\mathbf{p}-\mathbf{s}|^2} \di{A_{\mathbf{p}}} &
                \end{aligned}
                \label{eq:nc-model}
            \end{equation}
            
\noindent where $N_{\mathbf{o}}$ is the number of photons emitted in one pulse of light from $\mathbf{o}$, while $\rho$ denotes the albedo of the relay surface. The coefficients $A_{\mathbf{d}}$ and $A_{\mathbf{s}}$ are the active area of the detector and the area mapped on the relay surface, respectively. The unit vector $\bs{\omega}_{\mathbf{a}\to\mathbf{b}} = \frac{\mathbf{b}-\mathbf{a}}{|\mathbf{b}-\mathbf{a}|}$ represents the direction from the input argument $\mathbf{a}$ to $\mathbf{b}$. Adopting the same notation, $\mathbf{n}_{\mathbf{a}}$ represents the surface normal vector at the point $\mathbf{a}$. The Dirac delta function $\delta(\cdot)$ relates the time $t$ to the travel distance, while $c$ is the speed of light. The geometry term $\upsilon(\mathbf{p};\mathbf{l},\mathbf{s})$ is the visibility function of a hidden point $\mathbf{p}$ from an illumination spot $\mathbf{l}$ and a sensing patch $\mathbf{s}$. The function $f(\mathbf{p};\bs{\omega}_{\mathbf{l}\to\mathbf{p}},\bs{\omega}_{\mathbf{p}\to\mathbf{s}})$ describes the bidirectional reflectance distribution function (BRDF) of a point $\mathbf{p}$ with the incident and exitant directions $\bs{\omega}_{\mathbf{l}\to\mathbf{p}}$ and $\bs{\omega}_{\mathbf{p}\to\mathbf{s}}$. The integral $\int_{P}$ represents a summation of the photons that travel back at time $t$ from a small area $A_{\mathbf{p}}$ centered at the point $\mathbf{p}$ on the hidden object $P$.

For simplicity, Equation \eqref{eq:nc-model} neglects the volatility of light, e.g., diffraction and interference. We provide its complete derivation in Supplementary Information. $\tau(t;\mathbf{o},\mathbf{l},\mathbf{s},\mathbf{d})$ considers arbitrary combinations of illumination and detection points on the relay surface and records 5D transients, each of which is a histogram of the number of photons at time bins $t$. The peaks of the histogram indicate the arrival time of photons that travel back from the relay surface and from a hidden object in the scenario. $\tau(t;\mathbf{o},\mathbf{l},\mathbf{s},\mathbf{d})$ contains two portions: direct light paths in the LOS scene between the relay surface and the detector, and the indirect light paths in the NLOS scene between the relay surface and the hidden object. As these two portions are convolutional, we separate $\tau(t;\mathbf{o},\mathbf{l},\mathbf{s},\mathbf{d})$ into portions $\tau_{\text{NLOS}}(t;\mathbf{l},\mathbf{s})$ and $\tau_{\text{LOS}}(t;\mathbf{l}, \mathbf{s}, \mathbf{d})$, resulting in:

            \begin{equation}
                \begin{aligned}
                    \tau_{\text{NLOS}}(t;\mathbf{l},\mathbf{s}) 
                    &= 
                    \int_{P} \delta\rdb{|\mathbf{p}-\mathbf{l}|+|\mathbf{s}-\mathbf{p}| - ct} 
                    \upsilon(\mathbf{p};\mathbf{l},\mathbf{s}) \\ &\cdot 
                    f(\mathbf{p};\bs{\omega}_{\mathbf{l}\to\mathbf{p}},\bs{\omega}_{\mathbf{p}\to\mathbf{s}})
                    g(\mathbf{p}; \mathbf{l}, \mathbf{s})
                    \di{A_{\mathbf{p}}}
                \end{aligned}
                \label{eq:nc-model-NLOS}
            \end{equation}

\noindent and 

            \begin{equation}
                \begin{aligned}
                    \tau_{\text{LOS}}(t;\mathbf{l}, \mathbf{s}, \mathbf{d}) 
                    &= \Gamma(\mathbf{s}) \delta\rdb{|\mathbf{l}-\mathbf{o}|+|\mathbf{s}-\mathbf{d}| - ct}
                \end{aligned}
                \label{eq:nc-model-LOS}
            \end{equation}

\noindent where

             \begin{equation}
                g(\mathbf{p}; \mathbf{l}, \mathbf{s}) = 
                \frac{\rdb{\bs{\omega}_{\mathbf{p}\to\mathbf{l}} \cdot \mathbf{n}_{\mathbf{p}}} \rdb{\bs{\omega}_{\mathbf{l}\to\mathbf{p}} \cdot \mathbf{n}_{\mathbf{l}}} \rdb{\bs{\omega}_{\mathbf{s}\to\mathbf{p}} \cdot \mathbf{n}_{\mathbf{s}}} \rdb{\bs{\omega}_{\mathbf{p}\to\mathbf{s}} \cdot \mathbf{n}_{\mathbf{p}}}}{|\mathbf{l}-\mathbf{p}|^{2}|\mathbf{p}-\mathbf{s}|^2}
            \end{equation}
            
            \begin{equation}
                \Gamma(\mathbf{s}) = \frac{N_{\mathbf{o}}}{c}  \rdb{\frac{\rho}{\pi}}^{2} A_{\mathbf{d}} A_{\mathbf{s}} \frac{\rdb{\bs{\omega}_{\mathbf{s}\to\mathbf{d}} \cdot \mathbf{n}_{\mathbf{s}}} }{|\mathbf{s}-\mathbf{d}|^{2}}
                \label{eq:Gamma}
            \end{equation}
 
The function $g(\mathbf{p}; \mathbf{l}, \mathbf{s})$ represents attenuation effects dependent on the distance and shading effects due to the surface normals of a hidden point $\mathbf{p}$, $\mathbf{l}$, and $\mathbf{s}$. $\Gamma(\mathbf{s})$ models the intensity variation of the light after scattering off the relay surface, and restricts scanning regions of the detector. For NLOS reconstruction, the LOS portion is often gated or removed to obtain the NLOS portion $\tau_{\text{NLOS}}$ by assuming a virtual light source at $\mathbf{l}$ and a virtual detector at $\mathbf{s}$. $\tau_{\text{LOS}}$ contains information, e.g., depth and reflectance, on the relay surface and enables us to calibrate the hardware devices. The calibration procedures, however, are complicated because the distances $|\mathbf{l}-\mathbf{o}|$ and $|\mathbf{s}-\mathbf{d}|$ are difficult to determine from the measurements only. 
 
A confocal image formation model, first proposed by O'Toole et al. \cite{2018LCT}, collocates illumination and detection points, i.e., $\mathbf{l}=\mathbf{s}$ and $\mathbf{o}=\mathbf{d}$, and collects a 3D subset of the transient $\tau(t;\mathbf{o},\mathbf{l},\mathbf{s},\mathbf{d})$, as:
 
           \begin{equation}
                \begin{aligned}
                    \tau(t;\mathbf{o},\mathbf{s}) 
                    &= \Gamma(\mathbf{s})
                    \int_{P} \delta\rdb{2(|\mathbf{s}-\mathbf{o}|+|\mathbf{p}-\mathbf{s}|) - ct} \\ &\cdot \upsilon(\mathbf{p};\mathbf{s}) f(\mathbf{p};\bs{\omega}_{\mathbf{p}\to\mathbf{s}})
                    \frac{\rdb{\bs{\omega}_{\mathbf{p}\to\mathbf{s}} \cdot \mathbf{n}_{\mathbf{p}}}^2 \rdb{\bs{\omega}_{\mathbf{s}\to\mathbf{p}} \cdot \mathbf{n}_{\mathbf{s}}}^2}{|\mathbf{p}-\mathbf{s}|^{4}} \di{A_{\mathbf{p}}}
                \end{aligned}
                \label{eq:c-model}
            \end{equation}
 
 Similarly, we separate $\tau(t;\mathbf{o},\mathbf{s})$ into the LOS and NLOS portions of light paths as:

            \begin{equation}
                \begin{aligned}
                    \tau_{\text{NLOS}}(t;\mathbf{s}) 
                    &=
                    \int_{P} \delta\rdb{2|\mathbf{p}-\mathbf{s}| - ct} \upsilon(\mathbf{p};\mathbf{s}) f(\mathbf{p};\bs{\omega}_{\mathbf{p}\to\mathbf{s}})
                    g(\mathbf{p};\mathbf{s})
                    \di{A_{\mathbf{p}}}
                \end{aligned}
                \label{eq:model}
            \end{equation}
            
\noindent and
 
           \begin{equation}
                \begin{aligned}
                    \tau_{\text{LOS}}(t;\mathbf{s}) 
                    &=
                    \Gamma(\mathbf{s}) \delta\rdb{2|\mathbf{s}-\mathbf{o}| - ct}
                \end{aligned}
                \label{eq:los-model}
            \end{equation}
            
\noindent where

        \begin{equation}
            g(\mathbf{p}; \mathbf{s}) = 
            \frac{\rdb{\bs{\omega}_{\mathbf{p}\to\mathbf{s}} \cdot \mathbf{n}_{\mathbf{p}}}^2 \rdb{\bs{\omega}_{\mathbf{s}\to\mathbf{p}} \cdot \mathbf{n}_{\mathbf{s}}}^2}{|\mathbf{p}-\mathbf{s}|^{4}}
        \end{equation}

The confocal NLOS imaging model is highly simplified and has advantages in terms of system calibration because the distance between $\mathbf{s}$ and $\mathbf{o}$ and the coordinates of $\mathbf{s}$ are easily determined. 

\section{Adjustable NLOS imaging}

\begin{figure}
    \centering
    \hfill
    \subfloat[Light paths between the hardware devices]{
        \begin{minipage}{.54\linewidth}
            \centering
            \includegraphics[height=24ex]{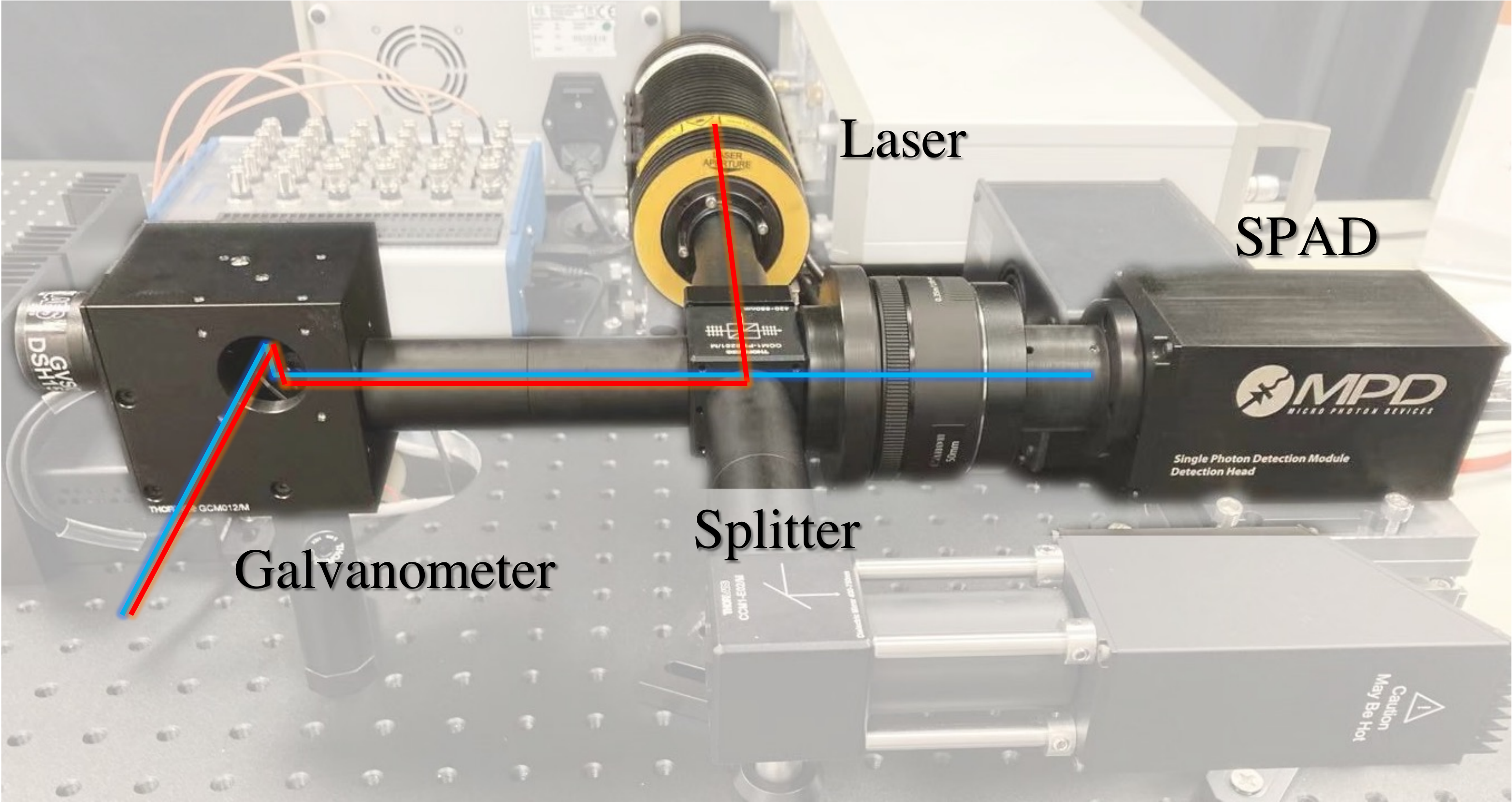}
        \end{minipage}
    \label{subfig:lightPath}}
    \subfloat[Opto-electrical design]{
        \begin{minipage}{.36\linewidth}
            \centering
            \includegraphics[height=24ex]{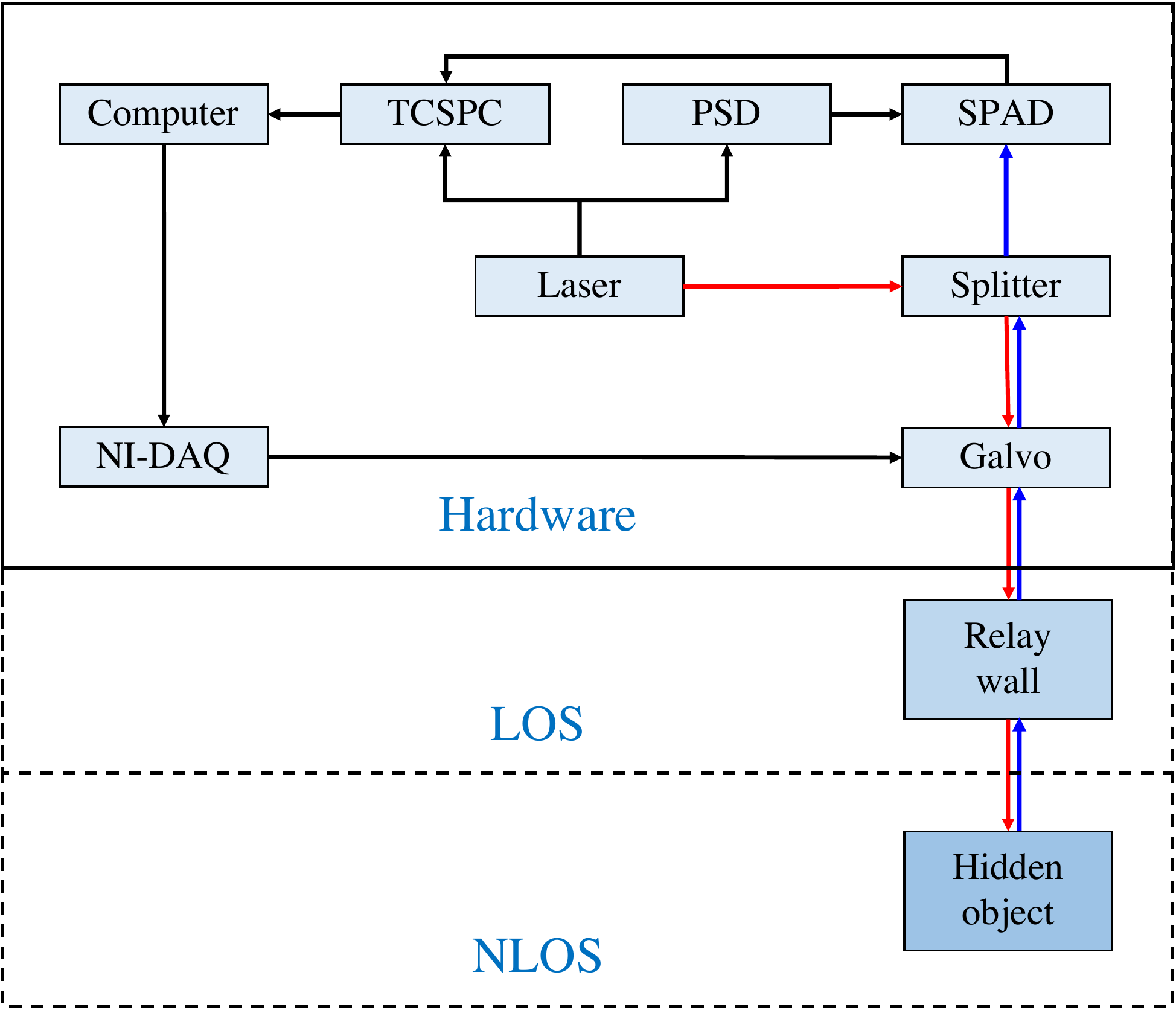}
        \end{minipage}
    \label{subfig:lightCircuit}}
    \hspace{\fill}~\\
    \subfloat[Procedural pipeline in software]{
        \begin{minipage}{.93\linewidth}
            \centering
            \includegraphics[width=.98\linewidth]{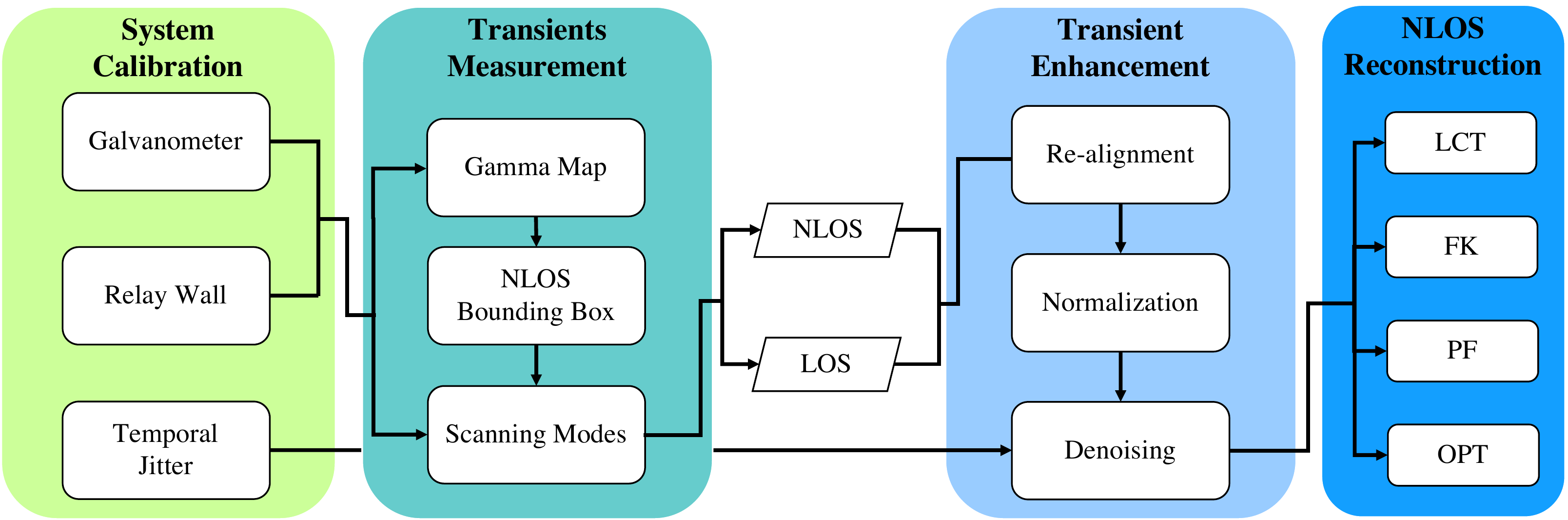}
        \end{minipage}
    \label{subfig:softwareProcess}}
    \caption{Overview of our capture system. (a) Light paths between a pulsed laser, a single-pixel SPAD, a beam splitter, and a 2D galvanometer; emitting light from the laser in red and returning light in blue. (b) The opto-electrical design of three modules: hardware module, LOS module, and NLOS module. (c) The pipeline of procedures in software. Note that the galvanometer calibration and the temporal jitter calculation are required only once, and any of the reconstruction algorithms is available as a default.}
    \label{fig:optodesign}
\end{figure}

Fig.~\ref{fig:optodesign} shows the overview of our adjustable NLOS imaging with the light paths and the opto-electrical design of the hardware devices. The capture system consists of three modules: hardware module, LOS module, and NLOS module. The hardware module includes hardware devices and their controllers. The LOS module relates the hardware module and the relay surface, and corresponds to the LOS portion of transients, which is exploited to define the relay surface, scanning patterns, and a measurable bounding box where hidden objects can be situated. The NLOS module models spherical light paths between the relay surface and the hidden objects, and contains the information on the hidden objects for reconstruction. Fig.~\ref{fig:optodesign}\subref{subfig:softwareProcess} illustrates the pipeline of the procedures in software that combines the applications provided by the manufacturers of the hardware devices, e.g., the TCSPC and the galvanometer, and operates the entire procedures with minimal human interventions. A software manipulation video is shown in Supplementary Information. 

\subsection{Gamma map}

In Equation \eqref{eq:Gamma}, we observe that in physics, $\Gamma(\mathbf{s})$ is related to the laser with $N_{\mathbf{o}}$, the detector with $A_{\mathbf{d}}$, the relay surface with $\frac{\rho}{\pi}$, and the relay surface with the detection points $\mathbf{s}$. It also relates the detector with the relay surface via the mapping from $A_{\mathbf{d}}$ to $A_{\mathbf{s}}$, and via the distance $|\mathbf{s}-\mathbf{d}|$. When the detector $\mathbf{d}$ is fixed, we can define $\Gamma(\mathbf{s})$ by determining the coordinates of detection points $\mathbf{s}$ on the relay surface. Based on the confocal NLOS imaging model, we calculate the coefficient of $\Gamma$ as the summation of the LOS portion $\tau_{\text{LOS}}(t;\mathbf{s})$ in Equation~\eqref{eq:los-model} at each detection point because the integral of $\delta$ is one. The coefficients of $\Gamma(\mathbf{s})$, or a \textit{Gamma} map, is defined for all detection points scanned on the relay surface. 

Fig.\ref{fig:scanRegion}\subref{subfig:Gamma} shows three \textit{Gamma} maps. We first scanned a small number of detection points to outline the region scannable by the galvanometer and calculate the \textit{Gamma} map (upper left) to adjust where the relay surface is situated with respect to the galvanometer. The \textit{Gamma} map (lower left) is then calculated to check the scanning region where the entire setting of the capture system is occluded by an object or its parts. After selecting a scanning pattern, we can calculate a \textit{Gamma} map (upper right) to verify the scanning region defined by a scanning pattern for NLOS measurements. The \textit{Gamma} map is also employed to normalize the transients because it models intensity variation of the NLOS measurements. 
The maximum intensity projection (MIP) map of LOS portions (lower right) appears similar to the corresponding \textit{Gamma} map, and can also be useful as the latter. The MIP map relies on the maximal value of the distribution function, rather than the integral, which may be less sensitive to the noise in transients measurements.

\subsection{System calibration}

The aim of calibration is to optimize our NLOS imaging by minimizing the difference between the transients that are measured from our setup and those that are theoretically computed based on the confocal NLOS imaging model. With the exception of hardware alignment, which needs human interventions to adjust light paths between hardware devices, the subsequent calibration procedures for the galvanometer, the relay surface, and the NLOS bounding box are iterative and can be operated online. The \textit{Gamma} maps are exploited to preview efficient scanning regions. \\[-1.5ex]

\noindent \textbf{Hardware alignment.} As shown in Fig.~\ref{fig:optodesign}, the laser and the SPAD are first aligned by observing the speckle of the laser beam that illuminates a surface, e.g., a white paper. We adjust the positions and angles of the laser and SPAD to maximize the intensity of the speckle. We then situate the beam splitter, the galvanometer, and the relay wall such that the speckle is clear at its focus on the wall. \\[-1.5ex]

\begin{figure}
    \centering
    \hspace{\fill}
    \subfloat[Coordinate systems and scanning regions]{
        \begin{minipage}{.36\linewidth}
            \centering
            \includegraphics[height=28ex]{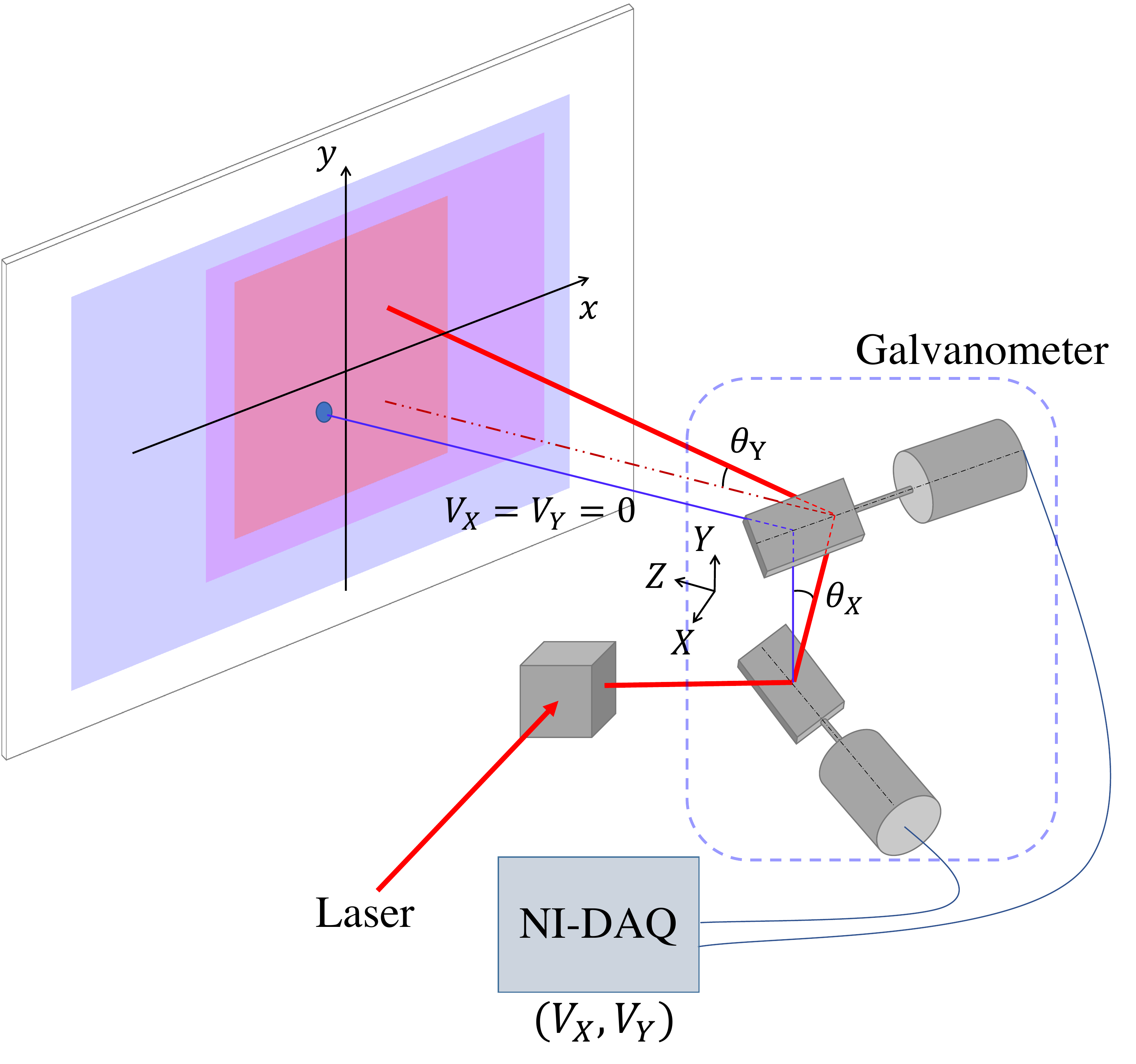}
        \end{minipage}
    \label{subfig:galvoSystem}}
    \subfloat[\textit{Gamma} and MIP maps]{
        \begin{minipage}[c][28ex]{.36\linewidth}
            \centering
            \includegraphics[height=25ex]{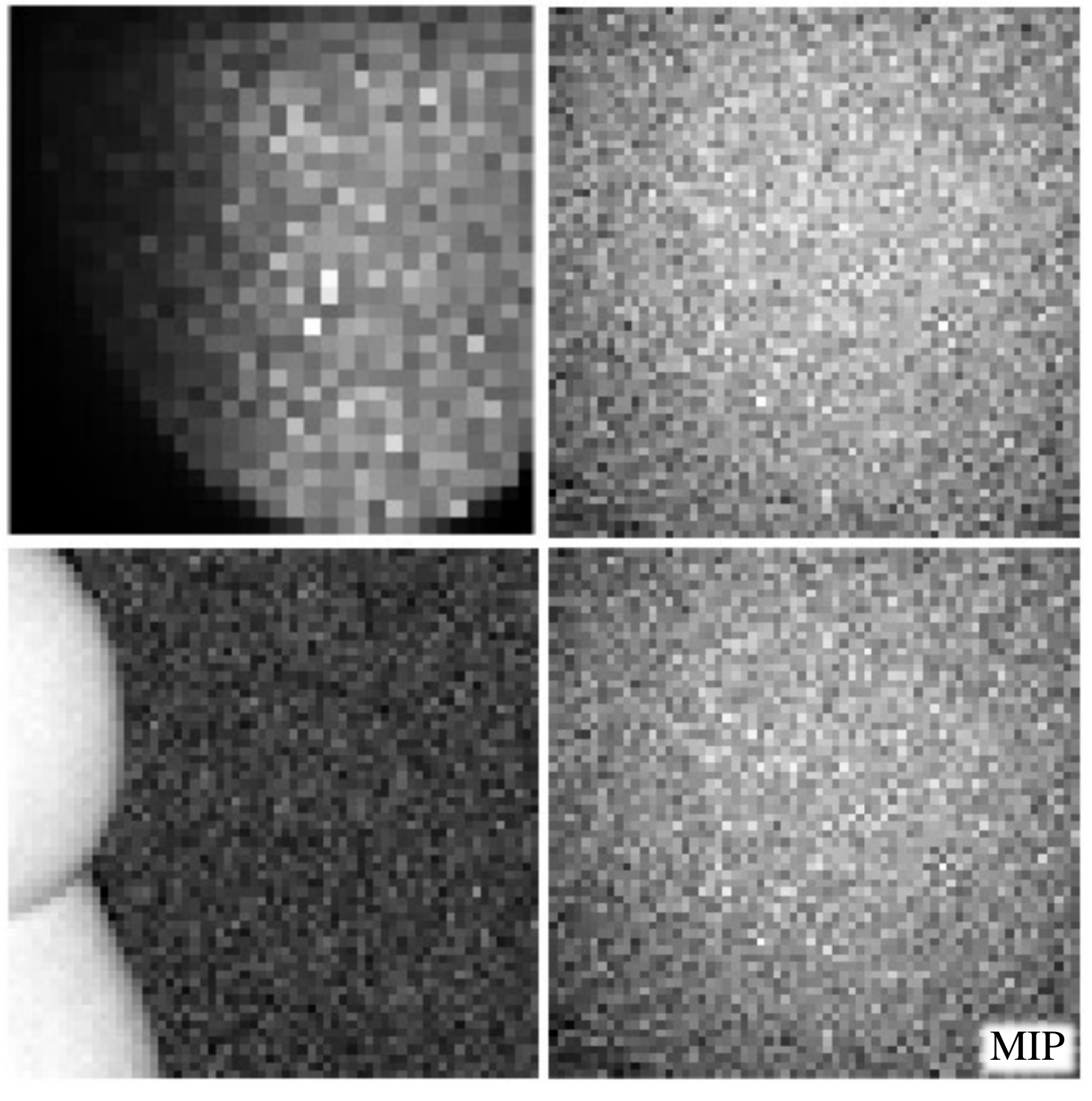}
        \end{minipage}
    \label{subfig:Gamma}}
    \hspace{\fill}~\\\hspace{-.02\linewidth}
    \subfloat[Arbitrary scanning pattern]{
        \begin{minipage}{.36\linewidth}
            \centering
            \includegraphics[width=.75\linewidth]{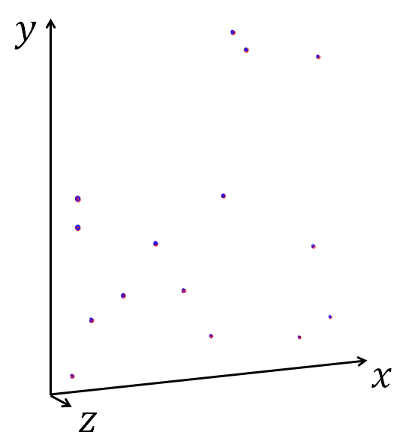}
        \end{minipage}
    \label{subfig:validationRandom}}\hspace{-.115\linewidth}
    \subfloat[Regular scanning pattern]{
        \begin{minipage}{.36\linewidth}
            \centering
            \includegraphics[width=.75\linewidth]{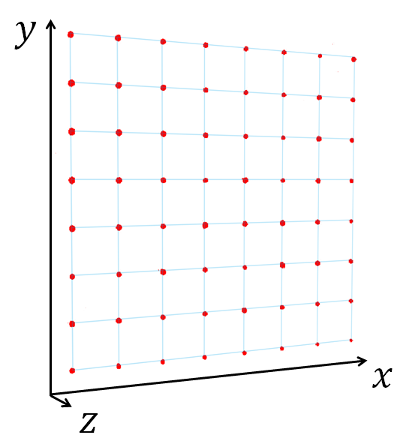}
        \end{minipage}
    \label{subfig:validationUniform}}\hspace{-.105\linewidth}
    \subfloat[Multi-circle scanning pattern]{
        \begin{minipage}{.36\linewidth}
            \centering
            \includegraphics[width=.75\linewidth]{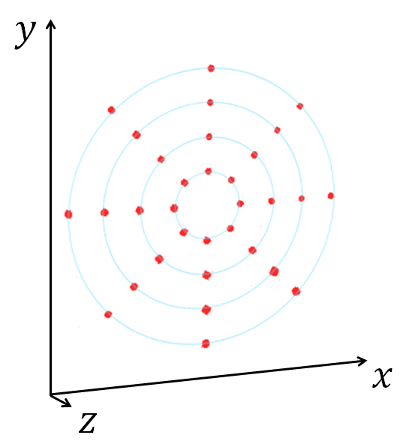}
        \end{minipage}
        \label{subfig:validationCircles}}
    \caption{Scanning regions and patterns. (a) Coordinate systems and scanning regions. Two coordinate systems: XYZ and xyz. Scanning regions: blue area defined by the galvanometer, magenta area limited by $\Gamma$, and pink area determined by scanning patterns. (b) \textit{Gamma} and MIP maps. The \textit{Gamma} maps of the scannable region by the galvanometer (upper left), of the efficient scanning region with occlusions (lower left), and of a scanned region defined by a scanning pattern (upper right), and a corresponding MIP map (lower right). (c-e) Scanning patterns: arbitrary pattern, regular grid pattern, and multi-circle pattern. The scanned points are in red; the desired detection points, the regular grid, and the concentric circles in blue, which are all post-processed. }
    \label{fig:scanRegion}
\end{figure}

\noindent \textbf{Galvanometer.} A dual-axis galvanometer supports optical scanning angles of about $\pm40\si{\degree}$, depending on several factors such as the laser beam diameter and the input voltage. In general, the galvanometer is coupled with a servo motor, which provides the feedback angles of the mirrors while scanning. We exploit the feedback between input voltages and angles to calibrate the galvanometer and to further determine the positions of scanning points. Fig.~\ref{fig:scanRegion}\subref{subfig:galvoSystem} illustrates the coordinate systems and scanning regions. Two Cartesian coordinate systems include $XYZ$ at the origin $\mathbf{o}$ where the laser beam is emitted into the free space toward the relay surface, and $xyz$ with the origin at the center of the detection region (in pink) on the relay surface, while $z=0$ when the relay surface is planar.


To calibrate the galvanometer, we assume that the scanning system is linear, and formulate the relationship between the optical scanning angles $\theta_{X}$ and $\theta_{Y}$ and the input voltages $V_{X}$ and $V_{Y}$, as:   
        
            \begin{equation}
                \begin{bmatrix}
                    \theta_{X} \\
                    \theta_{Y}
                \end{bmatrix}
                = 
                \begin{bmatrix}
                    \epsilon_{X} \\
                    \epsilon_{Y}
                \end{bmatrix}
                +
                \begin{bmatrix}
                    \beta_{XX} & \beta_{XY} \\
                    \beta_{YX} & \beta_{YY}
                \end{bmatrix}
                \begin{bmatrix}
                    V_{X} \\
                    V_{Y}
                \end{bmatrix}
                \label{eq:galvanometer_model}
            \end{equation}
            
\noindent where the initial angles $\epsilon_{X}$ and $\epsilon_{Y}$ are determined by the offset arrangement of the two mirrors prior to input voltages, and $\beta$ represents the coefficients of an angle with respect to the input voltages. The initial angles $\epsilon_{X}$ and $\epsilon_{Y}$ and the coefficients of $\beta$ may be offered by the manufacturers, whereas for precision, we collect $N$ groups of optical scanning angles $\theta_{n}$ along with input voltages $V_{n}$ in the given voltage range to calculate them. The coefficients of $\beta$ are first optimized using the multiple linear regression algorithm with a loss function $\mathcal{L}_{\text{Galvo}}$, as:    

            \begin{equation}
                \mathcal{L}_{\text{Galvo}}(\beta) = \frac{1}{N}\sum_{n}{|\theta_{n} - \beta V_{n}|^{2}}
                \label{eq:loss_galvo}
            \end{equation}

We then compute $\epsilon_{X}$ and $\epsilon_{Y}$ as the average error between the calculated and measured optical scanning angles. Finally, the input voltages for each group of scanning angles are calculated to control scanning points on arbitrary surfaces and to define the scannable region (blue area in Fig.~\ref{fig:scanRegion}\subref{subfig:galvoSystem}) of the galvanometer, which is unnecessarily rectangular. \\[-1.5ex]

\noindent \textbf{Relay surface.} The relay surface plays a critical role for an NLOS capture system. The LOS portion of transients, as in Equation~\eqref{eq:los-model}, can be exploited to recover the albedo and the orientation of the relay surface. Specifically, we extract the peak of the histogram $\tau_{\text{LOS}}(t;\mathbf{s})$ at each detection point and its corresponding $t$, and calculate the depth $\ell$ of the relay surface. We then employ the optical scanning angles $\theta_{X}$ and $\theta_{Y}$ of the galvanometer and the depth $\ell$ to estimate 3D coordinates of the detection points in $XYZ$, as:

        \begin{equation}
            \begin{cases}
                X = Z \tan{\theta_{X}} \\
                Y = Z \tan{\theta_{Y}} \\
                Z = \ell \rdb{1+\tan^{2}{\theta_{X}}+\tan^{2}{\theta_{Y}}}^{-1/2}
            \end{cases}
            \label{eq:coordinates-angledist}
        \end{equation}
        
The relay surface $W(W_{X},W_{Y},W_{Z})$ is considered to be planar and is formulated as: 

        \begin{equation}
            W: W_{X}X + W_{Y}X + W_{Z}Z + 1 = 0
            \label{eq:wall-model}
        \end{equation}
        
\noindent We notice that the albedo and the orientation of the relay  are considered in $\Gamma(\mathbf{s})$. Our calibration technique therefore does not require any additional devices or textured targets by using the \textit{Gamma} map. We thus recover the relay  when the loss function $\mathcal{L}_{W}(W_{X},W_{Y},W_{Z})$ is minimal, which is the root-mean-square error (RMSE) of distances from $N$ points to plane:

          \begin{equation}
            \mathcal{L}_{W}(W_{X},W_{Y},W_{Z}) = \sqrt{\frac{1}{N}\sum_{X,Y,Z}{\frac{|W_{X}X + W_{Y}X + W_{Z}Z + 1|^{2}}{W_{X}^{2}+W_{Y}^{2}+W_{Z}^{2}}}}
            \label{eq:wall-loss}
        \end{equation}
       
\noindent \textbf{NLOS bounding box.} Reconstruction accuracy depends on several factors of the capture system, including the geometric setting of the hardware module and the detection region on the relay . For efficient measurements, we define a bounding box to specify where the objects are situated in an NLOS scene.

Ahn et al.\cite{2019Ahn} have mentioned that the reconstructed shape should be within the orthogonal projection of the scanning region on the relay wall. In practice, we make a free space to allow for larger hidden objects by exploiting a relay wall whose orientation and position are adjustable. This setting is equivalent to adjusting the position of the entire setting of hardware devices. The orthogonal projection of the scanning region thus restricts a measurable bounding box of the NLOS scene with the maximal width and height of the scanning region. The minimal depth of the bounding box is approximated as:
        \begin{equation}
            z_{\min} = ct_{\text{delay}}
        \end{equation}
 
\noindent where $t_{\text{delay}}$ represents the delay between the arrival times of photons that travel back from the relay  and from the hidden objects. Since the distance between hidden objects and the relay  plays a significant role in the attenuation of photons, we consider it to be the maximal depth $z_{\max}$ of the bounding box. Here we assume that the hidden object is a perfectly diffuse white sphere, i.e., its BRDF $f(\mathbf{p})=\frac{1}{\pi}$, and that the signal is larger than a bias $b$, as:
        \begin{equation}
            \Gamma(\mathbf{s}) \int_{P}{\frac{(\bs{\omega}_{\mathbf{s}\to\mathbf{p}}\cdot\mathbf{n}_{\mathbf{s}})^{2}}{\pi|\mathbf{p}-\mathbf{s}|^{4}} \di{A}_{\mathbf{p}}} 
            = \Gamma(\mathbf{s}) \int_{0}^{2\pi}\int_{0}^{\pi/2}{\frac{\cos^{2}{\theta}}{\pi z_{\max}^{4}} \di{\theta}\sin{\theta}\di{\phi}} \ge b
        \end{equation}

\noindent when $\Gamma(\mathbf{s})$ is minimal. The bias $b$ includes the dark counts of the capture system and the ambient photon flux. $z_{\max}$ is then computed:
        \begin{equation}
            z_{\max} = \rdb{\frac{2\Gamma_{\text{min}}}{3b}}^{1/4}
        \end{equation}
        
\noindent where $\Gamma(\mathbf{s})$ limits the scanning region and the volume of the bounding box. Fig.~\ref{fig:scanRegion}\subref{subfig:galvoSystem} shows the initial scannable region of the galvanometer (in blue), the \textit{Gamma}-restricted scanning region (in magenta), and a user-defined detection region (in pink) on the relay . The bounding box of the NLOS scene helps us to estimate the size and position of a hidden object and to roughly predict the reconstruction quality by placing a textured target, e.g., a checkerboard, at different positions. 

\subsection{Scanning patterns}

The scanning process relates the hardware module to the relay  and the NLOS bounding box, and can also preview the effects of the system calibration. In the literature, illumination and detection points are usually distributed in a regular grid with evenly spaced points on the relay . In contrast, our capture system addresses user-defined scanning patterns by calculating the input voltages that correspond to the coordinates of each detection point on the relay . Specifically, we first re-parameterize Equation~\eqref{eq:wall-model} with $w(w_{X},w_{Y},w_{Z})$, as:
    
        \begin{equation}
            \begin{aligned}
                w &= \rdb{W_{X}^2+W_{Y}^2+W_{Z}^{2}}^{-1/2} \\
                w_{X} &= w \cdot W_{X} \\
                w_{Y} &= w \cdot W_{Y} \\
                w_{Z} &= w \cdot W_{Z}
                \label{eq:wall-parameters}
            \end{aligned}
        \end{equation}
        
In the efficient scanning region restricted by $\Gamma$, we define a scanning area (pink in Fig.~\ref{fig:scanRegion}\subref{subfig:galvoSystem}) with the origin $\mathbf{o_s}$ at its center on the relay . New sets of orthonormal basis $\hat{x},\hat{y},\hat{z}$ and $\hat{X},\hat{Y},\hat{Z}$ are constructed to normalize the surface normal of each detection point $(x,y,z)$, as:

            \begin{equation}
                \begin{cases}
                    \hat{x} = \frac{\frac{1}{w_{x}}\hat{X} - \frac{1}{w_{z}}\hat{Z}}{\sqrt{w_{x}^{-2}+w_{z}^{-2}}}\\
                    \hat{y} = \hat{z} \times \hat{x} \\
                    \hat{z} = w_{X}\hat{X} + w_{X}\hat{Y} + w_{Z}\hat{Z}
                \end{cases}
                \label{eq:wall-basis}
            \end{equation}

\noindent where $\hat{z}$ is the unit vector of the surface normal. Note that $z=0$ for a planar relay , and the coordinates of a point $\mathbf{p}$ on the hidden object $P$ can therefore be denoted with a value of $z$ such that the two coordinate systems for a detection point are transformed as:

        \begin{equation}
                \begin{cases}
                   \mathbf{s} = \mathbf{o}_{\mathbf{s}} + x\hat{x} + y\hat{y} + 0\hat{z} & (\text{in } xyz \text{ coordinate system})\\
                   \mathbf{s} = \mathbf{o} + X\hat{X} + Y\hat{Y} + Z\hat{Z} & (\text{in } XYZ \text{ coordinate system})
                \end{cases}
                \label{eq:coordinates-trans}
            \end{equation}
            
With Equations \eqref{eq:galvanometer_model} and \eqref{eq:wall-basis}, the input voltages of the galvanometer are determined by calculating the scanning angles from the coordinates of any detection point on the relay . To validate our scanning technique, we have tested three scanning patterns: arbitrary pattern, regular grid pattern, and multi-circle pattern.

For the arbitrary scanning pattern, we randomly raster-scan several points in different directions, and determine the coordinates of the detection points $\mathbf{s}$ and the equation of the relay . New coordinates $\mathbf{s}_{W}$ are then calculated for these points on the estimated relay  and their corresponding input voltages. Using the input voltages, the galvanometer is controlled to re-scan on the relay . Fig.~\ref{fig:scanRegion}\subref{subfig:validationRandom} demonstrates the two groups of detection points: desired detection points $\mathbf{s}$ are in blue, and the re-scanned points $\mathbf{s}_{W}$ in red.
  
        \begin{equation}
            \mathbf{s}_{W} = \mathbf{s} - (\hat{z} \cdot \mathbf{s})\hat{z}
        \end{equation}

We also define a regular scanning pattern with our system. $N \times N$ detection points are scanned in a region of $L \times L$, and the coordinates of the detection point $\mathbf{s}(i,j)$ are represented as: 

    \begin{equation}
        \mathbf{s}(i,j) = \mathbf{o}_{\mathbf{s}} - \frac{L}{2}\hat{x} + \frac{L}{2}\hat{y} + \frac{j-1}{N-1}L\hat{x} - \frac{i-1}{N-1}L\hat{y}
    \end{equation}

\noindent Fig.~\ref{fig:scanRegion}\subref{subfig:validationUniform} shows a regular scanning pattern with evenly spaced points (in red) on the relay . Note that the grid in blue is post-processed to identify the equidistant spaces between the scanned points. 

Inspired by a circular scanning pattern in \cite{2020ECCV}, we further present a multi-circle scanning pattern. In a scanning region with the radius $R$, we define $N_{r}$ concentric circles, and $N_{\phi}$ points on each circle. The coordinates of the detection points $\mathbf{s}(i,j)$ are then determined as:

    \begin{equation}
        \mathbf{s}(i,j) = \mathbf{o}_{\mathbf{s}} + \frac{i}{N_{r}}R\cos\rdb{\frac{\pi}{2}-(j-1)\frac{2\pi}{N_{\phi}}}\hat{x} + \frac{i}{N_{r}}R\sin\rdb{\frac{\pi}{2}-(j-1)\frac{2\pi}{N_{\phi}}}\hat{y}
    \end{equation}
    
The spaces between the circles and between the points on each circle may be equal or unequal. Similarly, we scan 32 sensing points on 4 circles, i.e., 8 points on each circle, and demonstrate them on additional concentric circles (in blue) as in Fig.~\ref{fig:scanRegion}\subref{subfig:validationCircles}. The results of the three scanning patterns in Fig.~\ref{fig:scanRegion}\subref{subfig:validationRandom} to \subref{subfig:validationCircles} show that the scanned points are in good agreement with the desired positions in either regular or irregular fashion, and either evenly or unevenly spaced.   

\subsection{Transient enhancement}

The signal-to-noise ratio of transients that are collected from the NLOS capture system is influenced by two major factors: properties of the SPAD including photon detection efficiency, afterpulsing, and pileup; and temporal jitter of the laser and the SPAD, which models uncertainty in the time-resolving mechanism. We first re-align the histogram (or transient) of each detection point such that the arrival time of the direct reflection from the relay  locates at the position of zero. Using the \textit{Gamma} map, we then normalize the transients $\tau(t;\mathbf{s})$ for higher quality.

While Hernandez et al.\cite{2017Quercus} have introduced a computational model for a SPAD, we opt for the approximation model in \cite{2017Matthew} to describe the probability of detecting individual photon events in a histogram bin as a Poisson distribution $\text{Pois}$. The bias $b$ is usually considered to be independent of time at a detection point and increases the background noise of the transient. The transients recorded with a SPAD, $\tau^{\text{SPAD}}(t;\mathbf{s})$, are formulated as:
 
             \begin{equation}
                \tau^{\text{SPAD}}(t;\mathbf{s}) = \text{Pois}\rdb{(\tau*j)(t;\mathbf{s})+b}
                \label{eq:spad_model}
            \end{equation}
            
\noindent where $j$ represents the temporal jitter of the system. Following \cite{2019Sun}, the temporal jitter typically yields a curve having two parts: a Gaussian peak and an exponential tail, as: 

            \begin{equation}
                \begin{aligned}
                    j(t; \mu, \sigma, \kappa_{0}, \kappa_{1}, \gamma) 
                    &= \text{Gaus}(t; \mu, \sigma) + \gamma \text{Exp}(t; \mu, \kappa_{0}, \kappa_{1}) \quad (t>0)
                \label{eq:jitter}
                \end{aligned}
            \end{equation}
            
\noindent where

            \begin{equation}
                \text{Gaus}(t; \mu, \sigma) = \exp\rdb{-\frac{(t-\mu)^{2}}{2\sigma^{2}}} 
            \end{equation}

            \begin{equation}
                \text{Exp}(t; \mu, \kappa_{0}, \kappa_{1}) = \frac{1}{\sqrt{t}} \exp\rdb{-\frac{(t-\mu)^{2}}{\kappa_{0}t}}\rdb{1 + \frac{t-\mu}{\kappa_{1}t}}
            \end{equation}

\noindent where $\mu, \sigma, \kappa_{0}$, and $\kappa_{1}$ are the coefficients of the temporal jitter, and $\gamma$ is the weight of the exponential term. The LOS transients $\tau_{\text{LOS}}^{\text{SPAD}}$ are exploited to calculate the temporal jitter $j$ of our capture system from the loss function $\mathcal{L}_{\text{SPAD}}$ as a cross-entropy loss:

            \begin{equation}
                \mathcal{L}_{\text{SPAD}}(\mu, \sigma, \kappa_{0}, \kappa_{1}, \gamma) = -\sum_{t,\mathbf{s}}{\left( \frac{\tau_{\text{LOS}}^{\text{SPAD}}(t;\mathbf{s})}{\Gamma(\mathbf{s})} \cdot \log{j(t-|\mathbf{s}|/c; \mu, \sigma, \kappa_{0}, \kappa_{1}, \gamma)} \right)}
            \end{equation}
            
We compute the temporal jitter at several detection points and consider the average value as the temporal jitter of our capture system. In Equation~\eqref{eq:spad_model}, we notice that the temporal jitter $j(t; \sigma, \gamma)$ dominates the measurement effect with a SPAD when decomposing the Poisson distribution from $\tau^{\text{SPAD}}$. The Wiener filter can thus be applied to denoise the transients $\tau^{\text{SPAD}}$:

            \begin{equation}
                \tilde{\tau}(\nu;\mathbf{s}) = \tilde{k}(\nu)\tilde{\tau}^{\text{SPAD}}(\nu;\mathbf{s})
            \end{equation}

\noindent where $\tilde{\tau}(\nu;\mathbf{s})$ and $\tilde{\tau}^{\text{SPAD}}$ represent the Fourier transform of $\tau(t;\mathbf{s})$ and $\tau^{\text{SPAD}}(t;\mathbf{s})$, while $\nu$ is the frequency. The kernel of the Wiener filter $\tilde{k}(\nu)$ in the frequency domain is computed from the Fourier transform of the temporal jitter, $\tilde{j}(\nu)$, and the signal-to-noise ratio $\eta$, as:  

            \begin{equation}
                \tilde{k}(\nu) = \frac{\rdb{\tilde{j}(\nu)}^{-1}}{1+\rdb{\eta|\tilde{j}(\nu)|^{2}}^{-1}}
            \end{equation}
            
O'Toole et al. \cite{2018LCT} embed the Wiener filter within their LCT algorithm, while Velten et al. \cite{2012Velten} use the Laplacian filter after the back-projection algorithm as a post-processing to mitigate noise. In contrast to LCT and FBP, we treat the denoising process prior to NLOS reconstruction such that we can opt for a wide variety of reconstruction algorithms. In common with the preceding studies, we ignore the effect of the Poisson distribution on denoising, while our optimization algorithm accounts for this effect on NLOS reconstruction. 

\subsection{NLOS Reconstruction}

Many algorithms have been proposed to reconstruct an NLOS scene from its transients. Volume-based algorithms, e.g., LCT\cite{2018LCT}, FK\cite{2019FK}, and PF\cite{2019Liu}, assume that the NLOS scene is represented as 3D voxels of a volume, and solve the inverse problem using FFT for reconstruction. These algorithms require a regular form of transients as input. Alternatively, optimization-based and learning-based algorithms (e.g., NeTF\cite{2021Shen}) support arbitrary forms of input transients, while the latter require tens of hours for training.  

We implement an optimization algorithm based on the confocal imaging model. In common with most existing methods, the optimization algorithm assumes that the hidden object is perfectly diffuse, and that there are no occlusions between the object and the relay surface, and no inter-reflection between the points on the hidden object. This implies that in Equation~\eqref{eq:model}, both $\upsilon(\mathbf{p};\mathbf{s})$ and $g(\mathbf{p}; \mathbf{s})$ are considered to be one, and the BRDF $f(\mathbf{p};\bs{\omega}_{\mathbf{p}\to\mathbf{s}})$ becomes view-independent $\mathbf{f}$. Equation~\eqref{eq:model} is thus simplified as a linear formulation:

    \begin{equation}
        \bs{\tau} = \Psi \mathbf{f}
    \end{equation}
    
\noindent where $\bs{\tau}$ is the discretized measurements and $\Psi$ a linearized measurement matrix. The albedo of the hidden object, $\mathbf{f}$, can be optimized by minimizing the differences between the theoretical $\Psi \mathbf{f}$ and the measured $\bs{\tau}$. We adopt the Poisson likelihood function to evaluate the similarity of the following observation function with total variation (TV) as a regularization term\cite{2021Ye,2012Harmany}: 
    
    \begin{equation}
        \mathcal{L}_{\text{NLOS}}\rdb{\mathbf{f}} =  -\ln\rdb{\prod_{i}{\frac{(\mathbf{e}^{\intercal}_{i} \Psi \mathbf{f})^{\tau_{i}}}{\tau_{i}!}} \exp({-\mathbf{e}^{\intercal}_{i} \Psi \mathbf{f}})} + \lambda ||\mathbf{f}||_{\text{TV}}
    \end{equation}
    
\noindent where $\mathbf{e}_i$ is the $i$th standard unit vector and $\tau_i$ the $i$th element of vector $\bs{\tau}$, while $\lambda$ is the weight of the TV. We adopt a gradient descent method to implement the optimization algorithm. 

\section{Evaluation}

To evaluate our capture system, we use transients that are theoretically simulated and really measured and reconstruct a number of NLOS scenes with the optimization algorithm (OPT) and state-of-the-art (SOTA) methods, including LCT\cite{2018LCT}, FK\cite{2019FK}, and PF\cite{2019Liu}.

\subsection{Simulation and measurement}

\textbf{Transient simulation.} We develop an algorithm to synthesize transients based on the confocal image formation model in Equation~\eqref{eq:model}. The illumination spot collocates with the detection point $\mathbf{s}$ on the relay , and the light scatters in a spherical wavefront toward a hidden object. We assume that at the detection point, an intensity map $L(u,v;\mathbf{s})$ and a depth map $D(u,v;\mathbf{s})$ are captured, where $u,v$ represents a corresponding pixel of the pair of maps. We then extract the depth and the intensity at each pixel, and compute the transient $\tau(t;\mathbf{s})$ at the time instant $t$. This process is repeated for all detection points, as outlined in Algorithm \ref{alg:simulation}.

    \begin{algorithm}
        \caption{Transient simulation}
        \label{alg:simulation}
        \begin{algorithmic}[1]
            \REQUIRE{depth maps $D(u,v;\mathbf{s})$, intensity maps $L(u,v;\mathbf{s})$, speed of light $c$}
            \ENSURE{transients $\tau(t;\mathbf{s})$}
            \STATE initialize $\tau(t;\mathbf{s})$
            \FORALL{sensing points $\mathbf{s}$}
                \FORALL{image pixels $(u,v)$}
                    \STATE $t = \text{round}(2*D(u,v)/c)$
                    \STATE $\tau(t;\mathbf{s}) = \tau(t;\mathbf{s}) + L(u,v)$
                \ENDFOR
            \ENDFOR
        \end{algorithmic}
    \end{algorithm}
    
\begin{figure}
    \centering
            \subfloat[Photograph of our capture system]{
                \begin{minipage}{.45\linewidth}
                    \centering
                    \includegraphics[height=36ex]{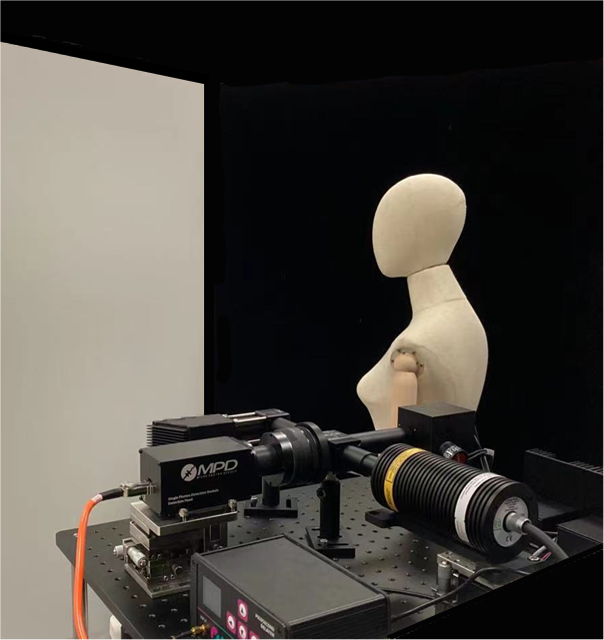}
                \end{minipage}
            \label{subfig:experimentSetup}}
            \subfloat[NLOS transients measured 
            and simulated
            ]{
                \begin{minipage}{.5\linewidth}
                    \centering 
                    \includegraphics[height=18ex]{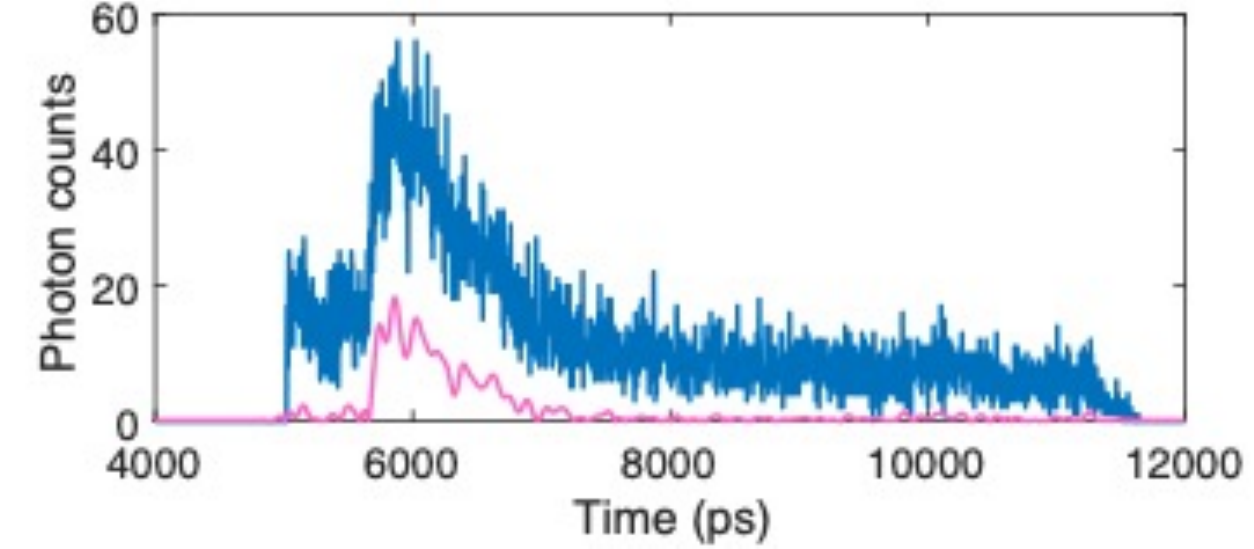}~\\
                    \includegraphics[height=18ex]{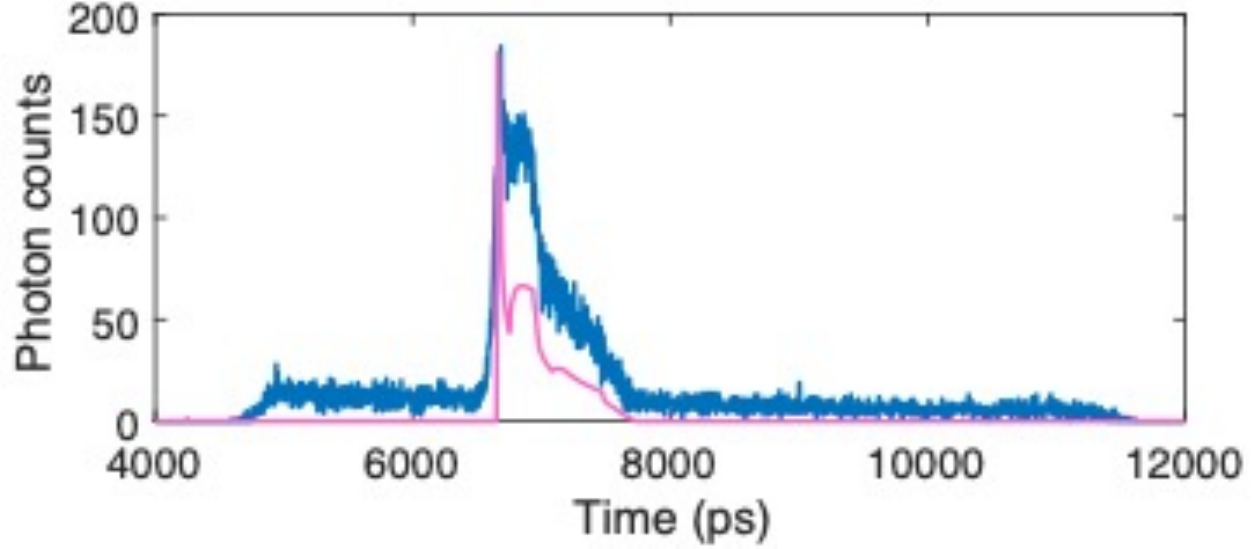}
                \end{minipage}
            \label{subfig:dataComparison}}
    \caption{Prototype and NLOS transients. (a) The prototype consists of a single-pixel SPAD, a pulsed laser, a 2D galvanometer, and a beam splitter. (b, upper) Transients measured, raw (blue) and denoised (red). (b, lower) Transients simulated with additional noise (blue) and without noise (red). Note that the transients are measured and simulated at the corresponding point of S-shape.}
    \label{fig:experimentSetup}
\end{figure}

We render transients $\tau(t;\mathbf{s})$ of an S-shape of \SI{0.6}{m} $\times$ \SI{0.6}{m} and a Whiteboard of \SI{0.6}{m} $\times$ \SI{0.4}{m} with \SI{4}{ps} temporal resolution and 64 $\times$ 64 spatial resolution. By considering the temporal jitter, the bias, and the Poisson distribution in Equation~ \eqref{eq:spad_model}, the transients with noise of the S-shape and the Whiteboard are also synthesized. Fig.~\ref{fig:experimentSetup}(b, lower) shows the simulated transients of S-shape with and without additional noise. \\[-1.5ex]

\noindent \textbf{Transient measurement.} Fig.~\ref{fig:experimentSetup}(a) shows a photograph of our capture system. We employ a fast-gated SPAD from Micro Photon Devices, which provides \SI{4}{ps} temporal resolution using a PicoHarp 300. An achromatic lens (Canon EF \SI{50}{mm} f/1.8) is exploited to focus the returning light onto the SPAD. A laser (PicoQuant LDH-IB-670-P) emits collimated light through a polarized beam splitter cube (Thorlabs VA5-PBS251). The light is then guided by a 2D galvanometer (Thorlabs GVS012) controlled by a data acquisition device. The SPAD records the indirect light from an NLOS scene by gating the direct light using a delayer (PicoQuant PSD-065-A-MOD). The temporal jitter of the entire system is approximately \SI{100}{ps} with a laser line filter (Thorlabs FL670-10). During the NLOS measurements, the laser power is set as \SI{1.54}{mW}, and delayer gate width is set as \SI{11}{ns}.

Our hardware devices are located \SI{1.0}{m} from the relay wall, a melamine white panel whose position and orientation can be adjusted for different sizes and shapes of NLOS scenes. We have measured the transients of a Whiteboard and an S-shape, whose settings are similar to their simulation. Fig.~\ref{fig:experimentSetup}(b, upper) shows measured transients, raw and denoised. In comparison with the simulated transients, the smaller peaks of measured transients are declined due to the properties of a SPAD, e.g., dark count rate and pileup. Other NLOS scenes include a Checkerboard, a Mannequin, and a Reso.board that is designed with stripes of different widths and lengths. All the objects measured are diffuse materials of paper, cotton or wood. Since the transients are captured automatically in our system, most experiments take \SI{30}{s} to record precise measurements at a single detection point and to transit to the next. The detailed parameters of experiments are shown in Table~\ref{tab:measurements}.

\subsection{Reconstruction results}

\begin{figure}
    \centering
    \includegraphics[width=.98\linewidth]{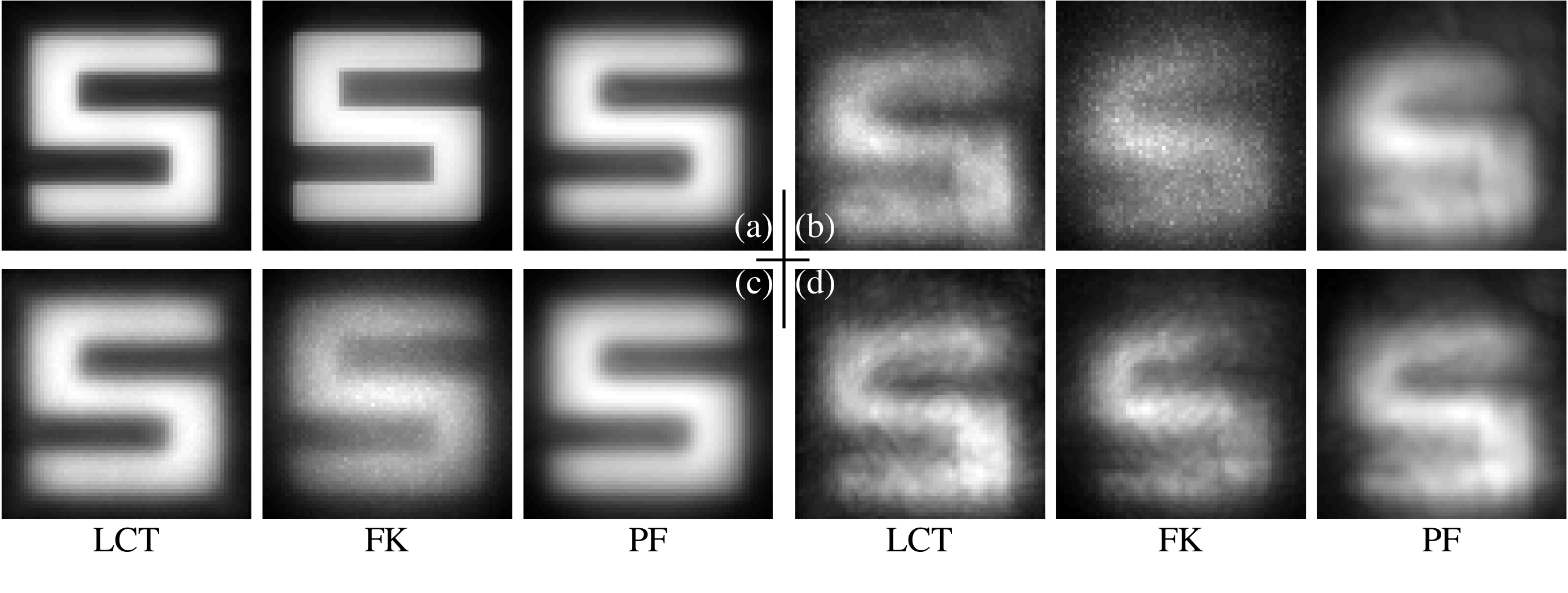}
    \caption{Evaluation of simulated and measured transients of S-shape. (a) and (c) are reconstructed using the simulated transients, without and with noise. (b) and (d) are reconstructed using the measured transients, raw and denoised.}
    \label{fig:sComparison}
\end{figure}

\begin{figure}
    \centering
    \includegraphics[width=.98\linewidth]{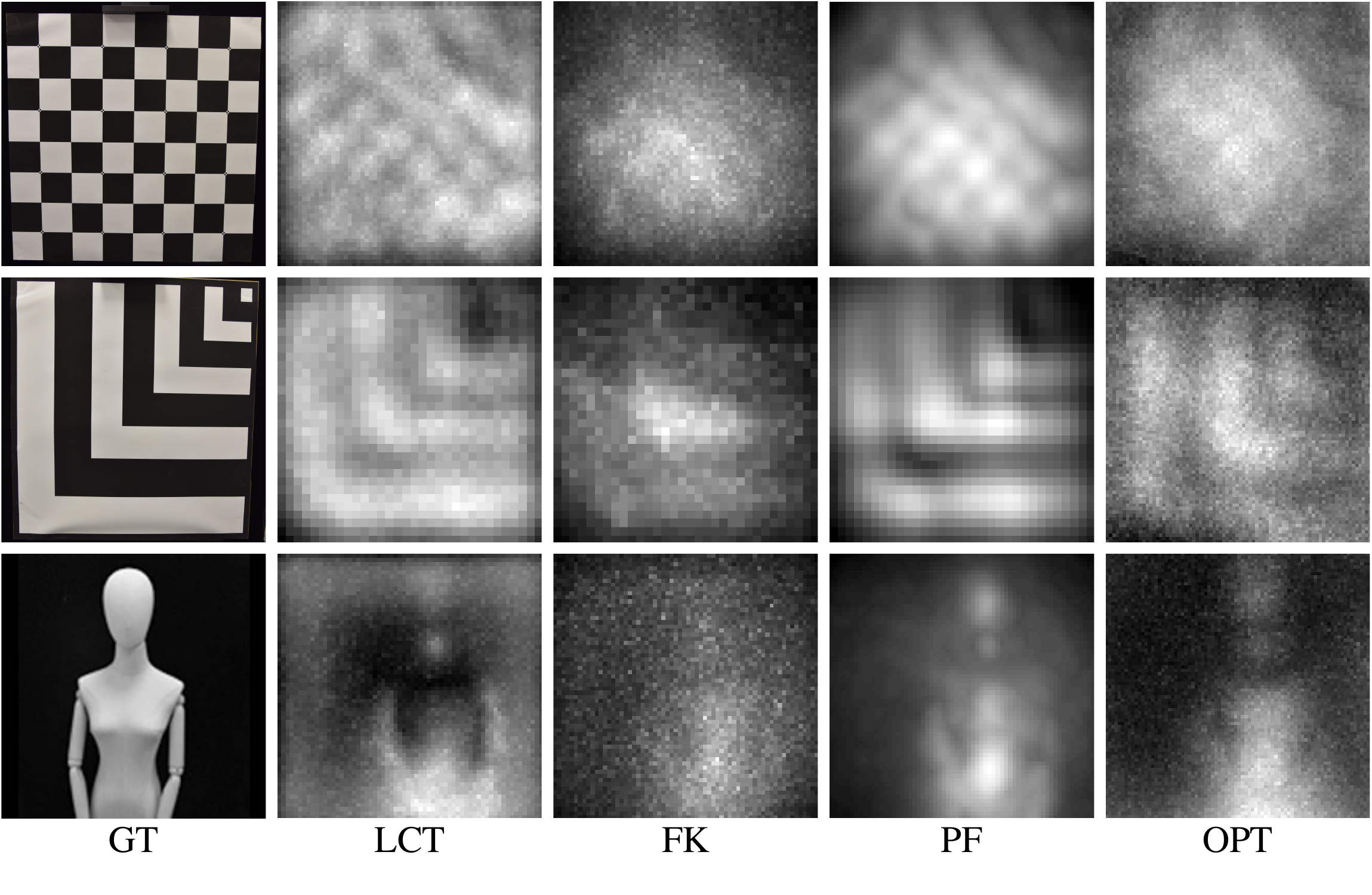}
    \caption{Evaluation of normalized transients in a regular grid. From top to bottom: Checkerboard, Reso.board, and Mannequin are reconstructed with the SOTA algorithms and OPT. The evaluation of denoised transients are shown in supplementary information}
    \label{fig:results}
\end{figure}

We have conducted experiments using the simulated and measured transients. The source codes with Matlab, including LCT, FK, and PF that are publicly available, run on a personal computer with an Intel i7-8750H CPU (2.2 GHz), 16 GB RAM, and a preliminary GPU 1050Ti. Fig.~\ref{fig:sComparison} shows the reconstruction of S-shape using the algorithms of LCT, FK, and PF. The transients of S-shape, raw or denoised, are measured in a regular grid and are simulated with and without noise. As expected, the volumes of S-shape are well reconstructed from the transients and our enhancement scheme can improve the reconstruction quality. We further carry out experiments of Checkerboard, Reso.board, and Mannequin using the normalized transients that are also measured in an equidistant grid. Fig.\ref{fig:results} demonstrates the volumes of the hidden objects reconstructed with the SOTA algorithms and OPT. The textured patterns of Checkerboard and Reso.board are shown in good agreement with the photographs or ground truth (GT), while the algorithms highly influence the reconstruction quality. The SOTA algorithms take of the order of seconds to reconstruct each NLOS scene, whereas our optimization algorithm takes several minutes for 1000 iterations. 


\begin{figure}
    \centering
    \includegraphics[width=.98\linewidth]{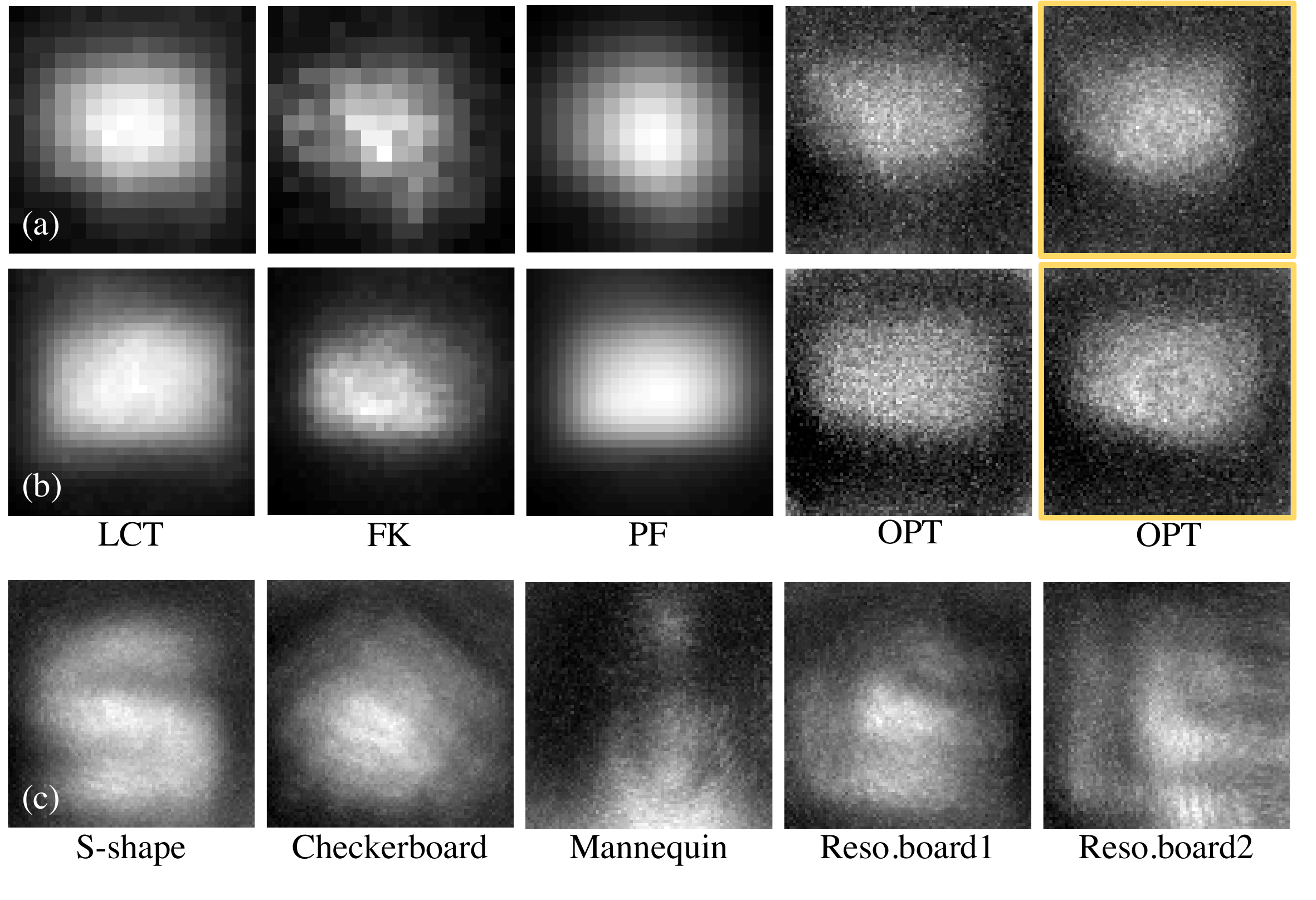}
    \caption{Evaluation of the transients measured in different scanning patterns. (a) and (b) Reconstruction results of Whiteboard with the SOTA algorithms and OPT, using transients in regular grid of 16 $\times$ 16 with \SI{15}{s} and 32 $\times$ 32 with \SI{30}{s}, while the two frames (yellow) are reconstructed with OPT using transients in a multi-circle pattern. (c) Reconstruction results with OPT of different objects. From left to right: Whiteboard, S-shape, Checkerboard, Mannequin, and Reso.board1 were measured in multi-circle pattern, while Reso.board2 was measured in an arbitrary pattern. }
    \label{fig:moreResults}
\end{figure}

In addition, we evaluate the reconstruction effects using our transients measured in different scanning patterns. Fig.\ref{fig:moreResults}(a) and (b) show the reconstruction results of Whiteboard with the SOTA algorithms and OPT. The transients of Whiteboard were measured in a regular grid of 16 $\times$ 16 with acquisition \SI{15}{s} and 32 $\times$ 32 with \SI{30}{s}, while the two frames in yellow were measured in a multi-circle pattern. We notice that the reconstruction quality with OPT using a smaller number of 
transients remains similar while the results with the SOTA algorithms are highly degraded. The Whiteboard measured in multi-circle pattern is slightly better reconstructed than in regular grid since denser detection points are distributed at the center of the NLOS scene. Fig.\ref{fig:moreResults}(c) shows the reconstruction results with OPT in more scanning patterns. Whiteboard, S-shape, Checkerboard, Mannequin, and Reso.board1 are measured in a multi-circle scanning pattern, and their volumes are sufficiently reconstructed. Reso.board2 is measured in an arbitrary pattern, i.e., coarse detection points in the lower left area and dense detection points in the upper right area, and is reconstructed with higher quality than Reso.board1 in multi-circle pattern. More experimental results with simulated and measured transients, raw and denoised, are shown in Supplementary Information.

\begin{table}
    \centering
    \caption{Experimental details of transient measurements. Three scanning patterns include regular grid (grid), multi-circle (circles), and arbitrary pattern.}
    \small
    \begin{tabular}{cccccc}
        \toprule
                Experiments & Whiteboard & S-shape & Checkerboard & Reso.board & Mannequin \\
        \midrule
            NLOS size & \SI{0.6}{m}$\times$\SI{0.4}{m} & \SI{0.6}{m}$\times$\SI{0.6}{m} & \SI{0.8}{m}$\times$\SI{0.8}{m} & \SI{0.8}{m}$\times$\SI{0.8}{m} & \SI{0.5}{m}$\times$\SI{0.6}{m} \\
            NLOS distance & \SI{0.8}{m} & \SI{0.8}{m} & \SI{0.8}{m} & \SI{0.8}{m} & \SI{0.8}{m} \\
            Detection region & \SI{0.8}{m}$\times$\SI{0.8}{m} & \SI{0.8}{m}$\times$\SI{0.8}{m} & \SI{0.8}{m}$\times$\SI{0.8}{m} & \SI{0.8}{m}$\times$\SI{0.8}{m} & \SI{0.8}{m}$\times$\SI{0.8}{m} \\
            Detection points & $16^{2}$/$32^{2}$ & $64^{2}$ & $64^{2}$ & $32^{2}$ & $64^{2}$ \\
            Scanning patterns & grid/circles & grid/circles & grid/circles & three patterns & grid/circles \\
            Exposure time & \SI{15}/\SI{30}{s} & \SI{30}{s} & \SI{30}{s} & \SI{30}{s} & \SI{30}{s} \\
        \bottomrule
    \end{tabular}
    \label{tab:measurements}
\end{table}

\section{Conclusion}

In this work, we have presented an adjustable NLOS imaging based on the confocal imaging model. Our capture system enables us to adjust scanning regions and patterns of detection points and to calibrate online the system components such as a relay surface and a galvanometer via \textit{Gamma} maps, scanning patterns, and reconstruction. In addition, our scanning technique allows users to define proper scanning patterns for precise measurements of different NLOS scenes, and 
the tailored reconstruction scheme can support arbitrary forms of input transients. Moreover, our software manipulates the procedures from system calibration to transient recording and enhancement to NLOS reconstruction except for manual adjustment of hardware devices. The experimental results for various hidden objects demonstrate that the measurements from our capture system are adequate and efficient. We believe that our work makes a significant step toward automation and convenience of NLOS acquisition and helps to facilitate NLOS imaging research.

Although our setup and techniques could be extended to support conventional NLOS imaging, estimating the coordinates of illumination and detection points may need additional devices or textured targets. The scanning patterns resemble how to align cameras and light sources in a light field imaging system and to determine rich information on the NLOS scenes in specific views. Adaptively selecting scanning points has the potential to unlock optimal numbers and positions of the illumination and detection points, resulting in minimal acquisition time and high-quality reconstruction. Our denoising scheme is preliminary and can be improved by accounting for more properties of a SPAD, e.g., Poisson distribution and pileup \cite{2019Gupta}, and by using a pulsed laser with even higher power. Further efforts and developments are thus needed for both NLOS imaging and NLOS reconstruction algorithms. \\

\noindent\textbf{Funding} This work was supported by Shanghai Science and Technology Program (21010502400) and NSFC programs (61976138, 61977047).

\noindent\textbf{Disclosures} The authors declare no conflicts of interest.

\noindent\textbf{Data availability} We will provide data underlying the results presented in this paper.

\bibliography{OnsiteNLOSImagingViaOnlineCalibrations}

\begin{thebibliography}{10}
\newcommand{\enquote}[1]{``#1''}

\bibitem{2020Faccio}
D.~Faccio, A.~Velten, and G.~Wetzstein, \enquote{Non-line-of-sight imaging,}
  {\protect\JournalTitle{Nature Reviews Physics}} \textbf{2}, 318--327 (2020).

\bibitem{2021Geng}
R.~Geng, Y.~Hu, and Y.~Chen, \enquote{Recent advances on non-line-of-sight
  imaging: Conventional physical models, deep learning, and new scenes,}
  {\protect\JournalTitle{arXiv preprint arXiv:2104.13807}}  (2021).

\bibitem{2012Gupta}
O.~Gupta, T.~Willwacher, A.~Velten, A.~Veeraraghavan, and R.~Raskar,
  \enquote{Reconstruction of hidden 3d shapes using diffuse reflections,}
  {\protect\JournalTitle{Optics express}} \textbf{20}, 19096--19108 (2012).

\bibitem{2015Mauro}
M.~Buttafava, J.~Zeman, A.~Tosi, K.~Eliceiri, and A.~Velten,
  \enquote{Non-line-of-sight imaging using a time-gated single photon avalanche
  diode,} {\protect\JournalTitle{Optics express}} \textbf{23}, 20997--21011
  (2015).

\bibitem{2018LCT}
M.~O'Toole, D.~B. Lindell, and G.~Wetzstein, \enquote{Confocal
  non-line-of-sight imaging based on the light-cone transform,}
  {\protect\JournalTitle{Nature}} \textbf{555}, 338--341 (2018).

\bibitem{2019FK}
D.~B. Lindell, G.~Wetzstein, and M.~O'Toole, \enquote{Wave-based
  non-line-of-sight imaging using fast fk migration,}
  {\protect\JournalTitle{ACM Transactions on Graphics (TOG)}} \textbf{38},
  1--13 (2019).

\bibitem{2019Liu}
X.~Liu, I.~Guill{\'e}n, M.~La~Manna, J.~H. Nam, S.~A. Reza, T.~H. Le,
  A.~Jarabo, D.~Gutierrez, and A.~Velten, \enquote{Non-line-of-sight imaging
  using phasor-field virtual wave optics,} {\protect\JournalTitle{Nature}}
  \textbf{572}, 620--623 (2019).

\bibitem{2020ECCV}
M.~Isogawa, D.~Chan, Y.~Yuan, K.~Kitani, and M.~O'Toole, \enquote{Efficient
  non-line-of-sight imaging from transient sinograms,} in \emph{European
  Conference on Computer Vision,}  (Springer, 2020), pp. 193--208.

\bibitem{2021Feng}
X.~Feng and L.~Gao, \enquote{Ultrafast light field tomography for snapshot
  transient and non-line-of-sight imaging,} {\protect\JournalTitle{Nature
  communications}} \textbf{12}, 1--9 (2021).

\bibitem{2021PNAS}
C.~Wu, J.~Liu, X.~Huang, Z.-P. Li, C.~Yu, J.-T. Ye, J.~Zhang, Q.~Zhang, X.~Dou,
  V.~K. Goyal \emph{et~al.}, \enquote{Non--line-of-sight imaging over 1.43 km,}
  {\protect\JournalTitle{Proceedings of the National Academy of Sciences}}
  \textbf{118} (2021).

\bibitem{2017Chan}
S.~Chan, R.~E. Warburton, G.~Gariepy, J.~Leach, and D.~Faccio,
  \enquote{Non-line-of-sight tracking of people at long range,}
  {\protect\JournalTitle{Optics express}} \textbf{25}, 10109--10117 (2017).

\bibitem{2021Metzler}
C.~A. Metzler, D.~B. Lindell, and G.~Wetzstein, \enquote{Keyhole imaging:
  Non-line-of-sight imaging and tracking of moving objects along a single
  optical path,} {\protect\JournalTitle{IEEE Transactions on Computational
  Imaging}} \textbf{7}, 1--12 (2020).

\bibitem{2012Velten}
A.~Velten, T.~Willwacher, O.~Gupta, A.~Veeraraghavan, M.~G. Bawendi, and
  R.~Raskar, \enquote{Recovering three-dimensional shape around a corner using
  ultrafast time-of-flight imaging,} {\protect\JournalTitle{Nature
  communications}} \textbf{3}, 1--8 (2012).

\bibitem{2020Ceiling}
J.~Rapp, C.~Saunders, J.~Tachella, J.~Murray-Bruce, Y.~Altmann, J.-Y.
  Tourneret, S.~McLaughlin, R.~M. Dawson, F.~N. Wong, and V.~K. Goyal,
  \enquote{Seeing around corners with edge-resolved transient imaging,}
  {\protect\JournalTitle{Nature communications}} \textbf{11}, 1--10 (2020).

\bibitem{2019Xin}
S.~Xin, S.~Nousias, K.~N. Kutulakos, A.~C. Sankaranarayanan, S.~G. Narasimhan,
  and I.~Gkioulekas, \enquote{A theory of fermat paths for non-line-of-sight
  shape reconstruction,} in \emph{Proceedings of the IEEE/CVF Conference on
  Computer Vision and Pattern Recognition,}  (2019), pp. 6800--6809.

\bibitem{2021Ye}
J.-T. Ye, X.~Huang, Z.-P. Li, and F.~Xu, \enquote{Compressed sensing for active
  non-line-of-sight imaging,} {\protect\JournalTitle{Optics Express}}
  \textbf{29}, 1749--1763 (2021).

\bibitem{2021Shen}
S.~Shen, Z.~Wang, P.~Liu, Z.~Pan, R.~Li, T.~Gao, S.~Li, and J.~Yu,
  \enquote{Non-line-of-sight imaging via neural transient fields,}
  {\protect\JournalTitle{IEEE Transactions on Pattern Analysis and Machine
  Intelligence}}  (2021).

\bibitem{2014Wu}
D.~Wu, G.~Wetzstein, C.~Barsi, T.~Willwacher, Q.~Dai, and R.~Raskar,
  \enquote{Ultra-fast lensless computational imaging through 5d frequency
  analysis of time-resolved light transport,}
  {\protect\JournalTitle{International journal of computer vision}}
  \textbf{110}, 128--140 (2014).

\bibitem{2013Heide}
F.~Heide, M.~B. Hullin, J.~Gregson, and W.~Heidrich, \enquote{Low-budget
  transient imaging using photonic mixer devices,} {\protect\JournalTitle{ACM
  Transactions on Graphics (ToG)}} \textbf{32}, 1--10 (2013).

\bibitem{2016Gariepy}
G.~Gariepy, F.~Tonolini, R.~Henderson, J.~Leach, and D.~Faccio,
  \enquote{Detection and tracking of moving objects hidden from view,}
  {\protect\JournalTitle{Nature Photonics}} \textbf{10}, 23--26 (2016).

\bibitem{2020NeRF}
B.~Mildenhall, P.~P. Srinivasan, M.~Tancik, J.~T. Barron, R.~Ramamoorthi, and
  R.~Ng, \enquote{Nerf: Representing scenes as neural radiance fields for view
  synthesis,} in \emph{European conference on computer vision,}  (Springer,
  2020), pp. 405--421.

\bibitem{2019Ahn}
B.~Ahn, A.~Dave, A.~Veeraraghavan, I.~Gkioulekas, and A.~C. Sankaranarayanan,
  \enquote{Convolutional approximations to the general non-line-of-sight
  imaging operator,} in \emph{Proceedings of the IEEE/CVF International
  Conference on Computer Vision,}  (2019), pp. 7889--7899.

\bibitem{2020Klein}
J.~Klein, M.~Laurenzis, M.~B. Hullin, and J.~Iseringhausen,
  \enquote{Calibration scheme for non-line-of-sight imaging setups,}
  {\protect\JournalTitle{Optics Express}} \textbf{28}, 28324--28342 (2020).

\bibitem{2015Ioannis}
M.~Laurenzis and A.~Velten, \enquote{Nonline-of-sight laser gated viewing of
  scattered photons,} {\protect\JournalTitle{Optical Engineering}} \textbf{53},
  023102 (2014).

\bibitem{2014Martin}
I.~Gkioulekas, A.~Levin, F.~Durand, and T.~Zickler, \enquote{Micron-scale light
  transport decomposition using interferometry,} {\protect\JournalTitle{ACM
  Transactions on Graphics (ToG)}} \textbf{34}, 1--14 (2015).

\bibitem{2011Kirmani}
A.~Kirmani, T.~Hutchison, J.~Davis, and R.~Raskar, \enquote{Looking around the
  corner using ultrafast transient imaging,}
  {\protect\JournalTitle{International journal of computer vision}}
  \textbf{95}, 13--28 (2011).

\bibitem{2019Lindell}
D.~B. Lindell, G.~Wetzstein, and V.~Koltun, \enquote{Acoustic non-line-of-sight
  imaging,} in \emph{Proceedings of the IEEE/CVF Conference on Computer Vision
  and Pattern Recognition,}  (2019), pp. 6780--6789.

\bibitem{2019Maeda}
T.~Maeda, Y.~Wang, R.~Raskar, and A.~Kadambi, \enquote{Thermal
  non-line-of-sight imaging,} in \emph{2019 IEEE International Conference on
  Computational Photography (ICCP),}  (IEEE, 2019), pp. 1--11.

\bibitem{2020Radar}
N.~Scheiner, F.~Kraus, F.~Wei, B.~Phan, F.~Mannan, N.~Appenrodt, W.~Ritter,
  J.~Dickmann, K.~Dietmayer, B.~Sick \emph{et~al.}, \enquote{Seeing around
  street corners: Non-line-of-sight detection and tracking in-the-wild using
  doppler radar,} in \emph{Proceedings of the IEEE/CVF Conference on Computer
  Vision and Pattern Recognition,}  (2020), pp. 2068--2077.

\bibitem{2017Jarabo}
A.~Jarabo, B.~Masia, J.~Marco, and D.~Gutierrez, \enquote{Recent advances in
  transient imaging: A computer graphics and vision perspective,}
  {\protect\JournalTitle{Visual Informatics}} \textbf{1}, 65--79 (2017).

\bibitem{2021Pei}
C.~Pei, A.~Zhang, Y.~Deng, F.~Xu, J.~Wu, U.~David, L.~Li, H.~Qiao, L.~Fang, and
  Q.~Dai, \enquote{Dynamic non-line-of-sight imaging system based on the
  optimization of point spread functions,} {\protect\JournalTitle{Optics
  Express}} \textbf{29}, 32349--32364 (2021).

\bibitem{2019Heide}
F.~Heide, M.~O'Toole, K.~Zang, D.~B. Lindell, S.~Diamond, and G.~Wetzstein,
  \enquote{Non-line-of-sight imaging with partial occluders and surface
  normals,} {\protect\JournalTitle{ACM Transactions on Graphics (ToG)}}
  \textbf{38}, 1--10 (2019).

\bibitem{2006Dutre}
P.~Dutre, K.~Bala, and P.~Bekaert, \emph{Advanced global illumination} (AK
  Peters/CRC Press, 2006), 2nd ed.

\bibitem{2017Quercus}
Q.~Hernandez, D.~Gutierrez, and A.~Jarabo, \enquote{A computational model of a
  single-photon avalanche diode sensor for transient imaging,}
  {\protect\JournalTitle{arXiv preprint arXiv:1703.02635}}  (2017).

\bibitem{2017Matthew}
M.~O'Toole, F.~Heide, D.~B. Lindell, K.~Zang, S.~Diamond, and G.~Wetzstein,
  \enquote{Reconstructing transient images from single-photon sensors,} in
  \emph{Proceedings of the IEEE Conference on Computer Vision and Pattern
  Recognition,}  (2017), pp. 1539--1547.

\bibitem{2019Sun}
F.~Sun, Y.~Xu, Z.~Wu, and J.~Zhang, \enquote{A simple analytic modeling method
  for spad timing jitter prediction,} {\protect\JournalTitle{IEEE Journal of
  the Electron Devices Society}} \textbf{7}, 261--267 (2019).

\bibitem{2012Harmany}
Z.~T. Harmany, R.~F. Marcia, and R.~M. Willett, \enquote{This is spiral-tap:
  Sparse poisson intensity reconstruction algorithms---theory and practice,}
  {\protect\JournalTitle{IEEE Transactions on Image Processing}} \textbf{21},
  1084--1096 (2011).

\bibitem{2019Gupta}
A.~Gupta, A.~Ingle, A.~Velten, and M.~Gupta, \enquote{Photon-flooded
  single-photon 3d cameras,} in \emph{Proceedings of the IEEE/CVF Conference on
  Computer Vision and Pattern Recognition,}  (2019), pp. 6770--6779.

\end{thebibliography}

\end{document}